\documentclass{aa}
\usepackage[varg]{txfonts}
\usepackage{bm}
\usepackage{amsmath}
\usepackage{cases}
\usepackage{ulem}
\usepackage{lineno}

\usepackage{graphicx}
\graphicspath{{../pdf/}{../jpeg/}}

\begin{document}

\title{Solar flares and Kelvin-Helmholtz instabilities: A parameter survey}
\author{W. Ruan \inst{\ref{inst1}} \and C. Xia \inst{\ref{inst1},\ref{inst2}} \and R. Keppens \inst{\ref{inst1}}}
\institute{Centre for mathematical Plasma Astrophysics, Department of Mathematics, KU Leuven, Celestijnenlaan 200B, B-3001 Leuven, Belgium \label{inst1}
\and School of Physics and Astronomy, Yunnan University, Kunming 650050, China \email{chun.xia@ynu.edu.cn} \label{inst2}}


\abstract
{Hard X-ray (HXR) sources are frequently observed near the top of solar flare loops, which are also bright in soft X-ray (SXR) and extreme ultraviolet (EUV) wavebands. We revisit a recent scenario proposed by \citet{Fang2016ApJ} to trigger loop-top turbulence in flaring loops, which can help explain variations seen in SXR and EUV brightenings and potentially impact and induce HXR emission. It is conjectured that evaporation flows from flare-impacted chromospheric footpoints interact with each other near the loop top and produce turbulence via the Kelvin-Helmholtz instability (KHI).}
{By performing a rigorous parameter survey, in which we vary the duration, total amount, and asymmetry of the energy deposition at both footpoints, we assess the relevance of the KHI in triggering and sustaining loop-top turbulence. We synthesize SXR and EUV emission and discuss the possibility of HXR emission through bremsstrahlung or inverse Compton processes,  which scatter SXR photons to HXR photons via the inverse Compton mechanism.} 
{We performed 2.5D numerical simulations in which the magnetohydrodynamic model incorporates a realistic photosphere to coronal stratification, parametrized heating, radiative losses, and field-aligned anisotropic thermal conduction.  We focus on the trigger of the KHI and the resulting turbulence, as well as identify various oscillatory patterns that appear in the evolutions.}
{We find that a M2.2-class related amount of energy should be deposited in less than one minute to trigger a KHI interaction. Slower deposition, or lesser energy ($<0.33 \times 10^{29}$ ergs) rather leads to mere loop-top compression sites bounded by shocks, without KHI development. Asymmetry in the footpoint deposition determines whether the KHI turbulent zone gets produced away from the apex, and asymmetric cases can show a slow-mode mediated, periodic displacement of the turbulent zone. Our reference simulation further demonstrates a clear $25 \,\,s$ periodicity in the declining phase of the SXR light curve, wherein compressional effects dominate.}
{When turbulence is produced in the loop apex, an index of -5/3 can be found in the spectra of velocity and magnetic field fluctuations. Typical values for M-class flares routinely show KHI development. The synthesized SXR light curve shows a clear periodic signal related to the sloshing motion of the vortex pattern created by the KHI.}



\maketitle

\section{Introduction}

Hard X-ray (HXR) emission in solar flares has been studied for decades. In most flares HXR emission is dominated by footpoint sources, but more and more observed flares show a more complex evolution \citep{Krucker2008A&ARv}. In addition to the footpoint sources, a loop-top source is reported in several flare observations (e.g. \citealp{Masuda1994Natur, Krucker2007ApJ, Su2013NatPh}). The mechanism of HXR sources is widely believed to be non-thermal thick target bremsstrahlung. According to the CSHKP flare model, part of the energy released by magnetic reconnection above the flare loop is used to accelerate particles \citep{Carmichael1964NASSP, Sturrock1966Natur, Hirayama1974SoPh, Kopp1976SoPh}. These accelerated electrons are guided by the magnetic field, and move from loop top to footpoints along the flare loops. Footpoint sources are generated when these energetic electrons collide with ambient ions in chromospheric footpoints and produce X-ray photons via bremsstrahlung.

Unlike foot-point sources, the loop-top source is usually not considered as a thick target. A coronal loop with a length of $10^9 \ \rm cm$ and a number density of $10^{10} \ \rm cm^{-3}$ is collisionally thin to electrons above $8 \ \rm keV$ \citep{Krucker2008A&ARv}, hence it seems difficult for the loop to stop energetic electrons with an energy of several ten $\ \rm keV$ efficiently enough to produce a loop-top HXR source. However, the efficiency of HXR photon generation increases rapidly when loop apex turbulence is taken into account. Turbulence is suggested to be an efficient trap for high energy electrons in the apex of flare loops \citep{Fang2016ApJ}. If turbulence exists in the apex, energetic electrons can effectively travel a much longer path before they leave the apex and the probability that the electrons collide or interact with ambient ions is effectively increased. Furthermore, more high energy electrons can participate in the bremsstrahlung to increase the intensity of loop-top HXR emission if the turbulence works as an efficient particle trap and accelerator. The efficiency of inverse Compton scattering (ICS) is improved by turbulence as well, since the probability that the electrons interact with soft X-ray (SXR) photons is also increased. Early on ICS was suggested,  in addition to bremsstrahlung
\citep{Korchak1967SvA}, as an alternative mechanism for HXR emission. The role of ICS can be more important than that of bremsstrahlung in HXR emission under special conditions, such as low plasma density, a hard spectrum of energetic electrons, and an anisotropic distribution of electrons \citep{Korchak1971SoPh, Chen2012ApJ}. \citet{Kontar2014ApJ} suggested the spectrum of loop-top electrons to be harder than expected in studying electron transport in flare loops, and they suggested transport is affected by turbulent pitch-angle scattering. A hard loop-top electron spectrum can imply that ICS  may play an important role in the HXR emission in the loop top.

The existence of turbulence in the apex of flare loops has been proved by multiple observations (e.g. \citealp{Antonucci1982SoPh, KontarPhysRevLett2017}). However, the origin of this turbulence needs further study. \citet{Fang2016ApJ} proposed a new probable turbulence source to interpret the generation of loop-top HXR emission. This new scenario suggests that Kelvin-Helmholtz instabilities (KHI) play an important role in the generation of turbulence in the apex. According to the CSHKP flare model, high speed evaporation flows are driven when high energy electrons deposit their energy into the chromospheric footpoints. The new scenario suggests that turbulence is produced via KHI when the flows go into the loop apex and interact with each other.  High speed evaporation flows with a speed of $200 - 500 \ \rm km/s$ are frequently reported \citep{Feldman1980ApJ, Antonucci1982SoPh, Shimizu1994ApJ, Milligan2009ApJ, Nitta2012SoPh, Tian2014ApJ, Tian2015ApJ}. \citet{Nitta2012SoPh} investigated 13 flare events, 3 of which were found to have high speed flows with a velocity about $500 \ \rm km/s$. Evaporating upflows with speeds higher than $500 \ \rm km/s$ seem relatively rare and only a few cases are reported (e.g. \citealp{Bentley1994ApJ, Tomczak1997A&A, Liu2006ApJ}). While the evaporation flows have a speed of several hundred kilometers per second, the speed of the flows is likely comparable to the local Alfv\'en speed. Therefore, it is reasonable to consider the contribution of KHI. \citet{Fang2016ApJ} proved that KHI can be triggered and turbulence can be produced in the apex of flare loops. In their observational study, \citet{Antonucci1982SoPh} also found a link between high speed evaporation flows and turbulence. The turbulence in hot plasma increases and can reach a maximum turbulent velocity of $122 \ \rm km/s$ when a high speed upflow with a speed of $300 - 400 \ \rm km/s$ has been observed, while the mean turbulent velocities is of the order of $75 \ \rm km/s$ in the preflare active region. We suggest that this increase of turbulent velocity is likely caused by KHI.

When the energetic particles deposit their energy into chromospheric footpoints, the energy is unlikely to be equally distributed to two footpoints. \citet{Aschwanden1999ApJ} found that the HXR emissions of two footpoints tend to be asymmetric in their survey, which indicates the asymmetry of energy deposition. The asymmetry of energy deposition is suggested to be determined by the asymmetry of the magnetic field. Figures 1 and 4 in \citet{Krucker2008A&ARv} also indicate a result of asymmetric energy deposition, in which one footpoint is closer to the loop-top HXR or SXR sources than the other. The asymmetry of energy deposition may influence the parameters of evaporation flows and the evolution of the flare loop.  In our numerical study of KHI triggering, asymmetric energy deposition strategies are adopted to make our simulations comparable with the observations.

We aim to continue the study of \citet{Fang2016ApJ} on the trigger of KHI and the generation of turbulence in flare loops. The set-up of our simulation is introduced in section 2. Energy is deposited into chromospheric footpoints of a coronal loop that drives evaporation flows. The interaction of the flows, trigger of KHI, and generation of turbulence are investigated in section 3. The thermodynamics and radiative evolution of the flare loop and energy cascade process of turbulence are studied with a high spatial resolution case. In addition, EUV and SXR images of the flare loop are synthesized in this section, and then the role of bremstrahlung versus the inverse Compton process in the generation of loop-top HXR sources is investigated. A parameter survey is performed to investigate important factors that influence the trigger of KHI and the generation of turbulence in section 4. We summarize our work in section 5.

\section{Numerical set-up}
We numerically solved the magnetohydrodynamics (MHD) equations with the MPI-parallelized Adaptive Mesh Refinement Versatile Advection Code \textit{MPI-AMRVAC} \citep{Keppens2012JCoPh, Porth2014ApJS, Xia2018ApJS}. The governing equations are written as follows:
\begin{eqnarray}
\frac{\partial \rho}{\partial t} + \nabla \cdot (\rho \bm{v}) & = & 0, \\ 
\frac{\partial \rho \bm{v}}{\partial t} + \nabla \cdot (\rho \bm{v} \bm{v} + p_{\rm{tot}} \bm{I} - \frac{\bm{BB}}{\mu_0}) & = & \rho \bm{g}, \\
\frac{\partial e}{\partial t} + \nabla \cdot (e \bm{v} + p_{\rm{tot}} \bm{v} - \frac{\bm{BB}}{\mu_0} \cdot \bm{v}) & = & \rho \bm{g} \cdot \bm{v} +
\nabla \cdot (\bm{\kappa} \cdot \nabla T) \nonumber \\
& &  - Q + H, \label{q-en} \\
\frac{\partial \bm{B}}{\partial t} + \nabla \cdot (\bm{v B} - \bm{B v}) & = & 0,
\end{eqnarray}
where $\bm{I}$ is the unit tensor and $\bm{\kappa}$ is the thermal conductivity tensor, and $\rho$, $\bm{v}$, $T,$ and $\bm{B}$ are density, velocity, temperature, and magnetic field, respectively. The total energy density is given by 
\begin{equation}
e = \frac{p}{\gamma-1} + \frac{\rho v^2}{2} + \frac{B^2}{2 \mu_0},
\end{equation}
and the total pressure is given by 
\begin{equation}
p_{\rm{tot}} = p + \frac{B^2}{2 \mu_0}.
\end{equation}
The terms $Q$ and $H$ in the energy equation (\ref{q-en}) indicate optically thin radiative cooling and parametrized heating, respectively. 

We solved the MHD equations on a Cartesian box with a domain of $-40\ \textrm{Mm} \leq x \leq 40\ \textrm{Mm}$ and $0\ \textrm{Mm} \leq y \leq 50\ \textrm{Mm}$, but vector quantities are 2.5D (i.e. they have three components). We adopted an abundance ratio of $\rm{He/H = 0.1}$ in the calculations of density and pressure, where the plasma is assumed to be fully ionized. The gravitational acceleration is given by $\bm{g} = g_0 R_{\rm{s}}^2 / (R_{\rm{s}} + y)^2 \bm{\hat{y}}$, where $g_0 = -274\ \rm m\ s^{-2}$ and $R_{\rm{s}}$ is the solar radius. The thermal conductivity tensor is expressed as $\bm{\kappa} = \kappa_{\parallel} \bm{\hat{b}} \bm{\hat{b}}$ where $\bm{\hat{b}}=\bm{B}/B$, which indicates that only thermal conduction along magnetic field lines was considered. The parallel thermal conductivity is given by $\kappa_{\parallel} = 8 \times 10^{-7} T^{5/2} \ \rm{erg\ cm^{-1}\ s^{-1}\ K^{-7/2}}$. We adopted a maximum heat flux $F_{\rm{sat}} = 5 \phi \rho c_{\rm{s}}^3$ to include the saturation effect of thermal conduction \citep{Cowie1977ApJ}, where $\phi = 1$ and $c_s$ denotes the local acoustic speed.  The radiative cooling function is expressed as $Q = n_{\rm{e}}^2 \Lambda (T)$, where $n_{\rm{e}}$ is the number density of electrons and $\Lambda (T)$ is a cooling curve supplied by \citet{Colgan2008ApJ}. This curve provides the radiative loss of hot and optically thin plasma; the emission from plasma with a temperature lower than $10,000 \ \rm K$ was not considered. 

The initial conditions and boundary conditions are as follows. The lower boundary is located in the photosphere and the upper boundary is located in the corona. The initial distribution of temperature consists of two parts: the VAL-C temperature profile  \citep{Vernazza1981ApJS} is employed in the region between the lower boundary and $h_{\rm{tra}} =2.543 \ \rm Mm$, and the temperature distribution in the region above $h_{\rm{tra}}$ can be written as
\begin{equation}
T(y) = [3.5 F_{\rm{c}} (y - h_{\rm{tra}}) / \kappa + T_{\rm{tra}}^{7/2}]^{2/7},
\end{equation}
where $F_{\rm{c}} = 2 \times 10^5 \ \rm erg \ cm^{-2} \ s^{-1}$, $\kappa = 8 \times 10^{-7} T^{5/2} \ \rm {erg\ cm^{-1}\ s^{-1}\ K^{-7/2}}$,  $T_{\rm{tra}} = 0.447 \ \rm MK$. 
There is a ghost zone below the lower boundary ($y = 0$) and the thickness of the ghost zone is $2 \ \rm Mm$. We obtained the temperature profile in the ghost zone from a linear extrapolation of the temperature profile above the lower boundary. The plasma density at the bottom of the ghost zone was set to be $n_{\rm{e}} = 7.1 \times 10^{14} \ \rm cm^{-3}$. The initial plasma density in the ghost zone and the simulation region was derived from hydrostatic equilibrium $\partial p / \partial h = \rho g$. As a result, the number density at $y = 0$ is $n_{\rm{e}} = 2.15 \times 10^{14} \ \rm cm^{-3}$.
Each boundary contains two layer of ghost cells. For the left and right boundary, $\rho$, $e$, $v_y$, and $B_y$ use a symmetric boundary, while $v_x$, $v_z$, $B_x$, and $B_z$ employ an anti-symmetric boundary. For the upper and bottom boundaries, the velocity components also employ an anti-symmetric boundary. In the lower boundary ghost cells, $\rho$ and $e$ and $\bm{B}$ were fixed at their initial values. For ghost cells of the upper boundary, $\rho$ and $e$ were calculated according to the gravity stratification, where the value of temperature was obtained from extrapolation. The value $\bm{B}$ in the upper boundary was extrapolated assuming zero normal gradient. We imposed zero velocities at the bottom and upper boundary surface. We used the same boundary conditions for all of the simulations.
We used adaptive mesh refinement  to refine the mesh and increase the spatial resolution based on the spatial smoothness of density and magnetic field. The initial simulation box contains $128 \times 80$ cells in a domain of $\rm 80\ Mm \times 50\ Mm$, and the maximum refinement level is set to five or six. A high spatial resolution of $19.5\ \rm km$ is achieved inside and near the flare loop, which ensured that we could simulate the triggering of KHI and evolution of turbulence. 

\begin{figure}[htbp]
\begin{center}
\includegraphics[width=0.8\linewidth]{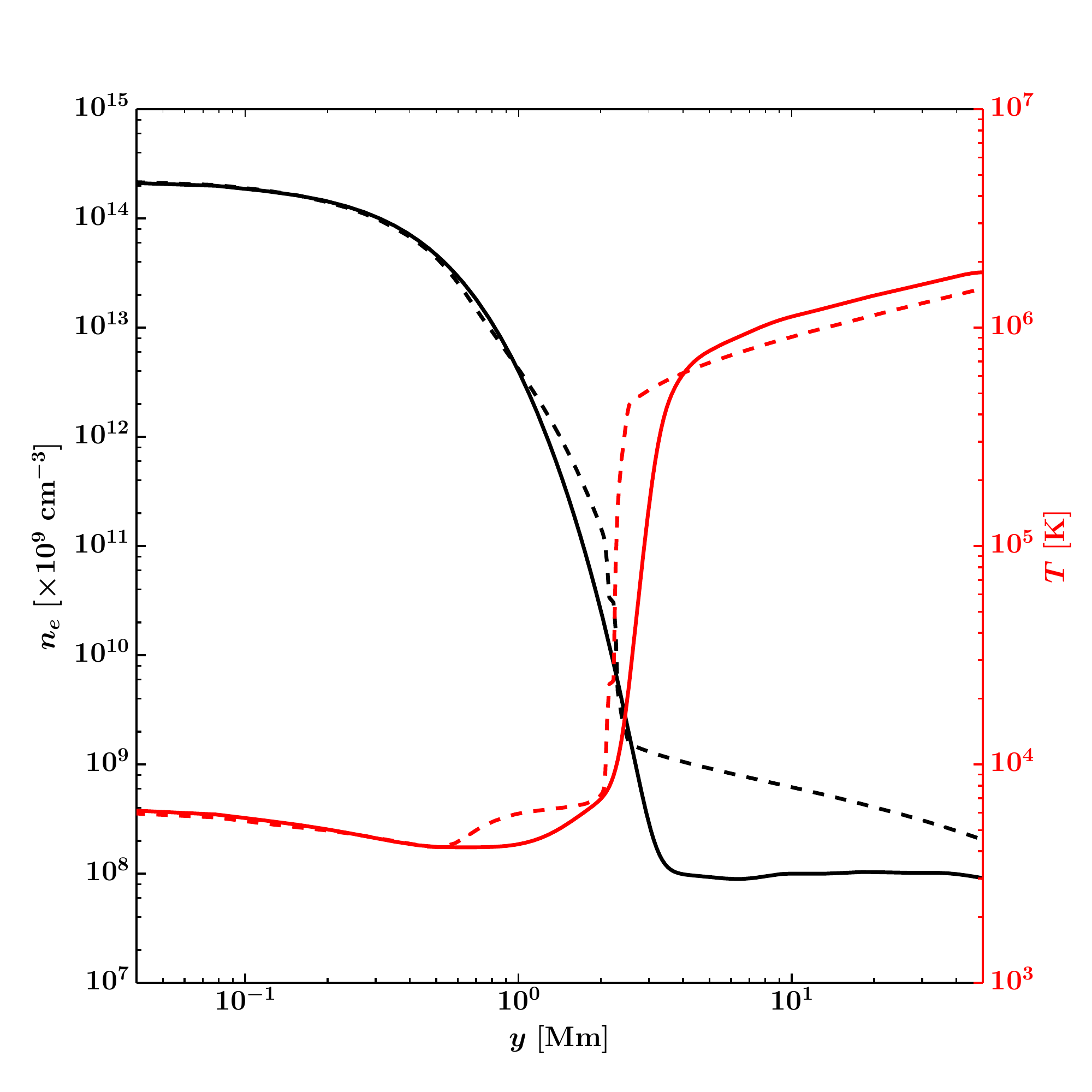}
\caption{Height profiles of temperature and density before relaxation (dashed lines) and after relaxation (solid lines).}
\label{rhoT_profile}
\end{center}
\end{figure}

A background heating is employed to heat the corona, and is given by
\begin{equation}
H_0 = c_0 \exp (- \frac{y}{\lambda_0}),
\end{equation}
where $c_0 = 10^{-4} \ \rm erg\ cm^{-3} \ s^{-1}$ and $\lambda_0 = 80 \ \rm Mm$. We adopted a force-free magnetic arcade as an initial magnetic configuration,
\begin{eqnarray}
B_x & = & -B_0 \cos (\frac{\pi x}{L_0}) \sin \theta_0 \exp (- \frac{\pi y \sin \theta_0}{L_0}), \\
B_y & = & B_0 \sin (\frac{\pi x}{L_0}) \exp (- \frac{\pi y \sin \theta_0}{L_0}), \\
B_z & = & -B_0 \cos (\frac{\pi x}{L_0}) \cos \theta_0 \exp (- \frac{\pi y \sin \theta_0}{L_0}),  
\end{eqnarray}
where $\theta_0 = 30^{\circ}$ is the angle between the apex of the magnetic loops and the neutral line, $L_0 = 80 \ \rm Mm$ is the horizontal size of the simulation box, and $B_0 = 80 \ \rm G$ is the strength of magnetic field at the bottom. The resulting magnetic field strength at the loop apex is about $50 \ \rm G$.
We performed relaxation with this initial condition and background heating to obtain a stable plasma environment during which the precise thermodynamic  balance gets adjusted. When the relaxation is finished, the maximum velocity in the simulation box is smaller than $4\ \rm km~s^{-1}$. The initial and the relaxed-state profiles of density and temperature at $x = 0$ are shown in Figure \ref{rhoT_profile}.

After the relaxation, we reset the time to zero and deposited flare energy into the chromospheric footpoints of a loop to trigger evaporation flows. According to the CSHKP flare model, high energy electron flows generated in the reconnection above the loop top move along the magnetic field, heat the lower atmosphere, and trigger the evaporation flows. The observation from Ramaty High Energy Solar Spectroscopic Imager (RHESSI) suggests that the high energy electron flows have a double power law spectrum  \citep{Holman2003ApJ}. The high energy particles are found to deposit most of their energy in the upper chromosphere owing to Coulomb collisions \citep{Allred2005ApJ}. This effect is incorporated and represented by an energy source $H_1$ added in the upper chromosphere in the flare loops where the initial magnetic loops fulfil the following inequality:
\begin{eqnarray}
&& A_z(24\ \textrm{Mm},0) < A_z(x,y) < A_z(23\ \textrm{Mm}, 0), \label{q-aset}\\
&& A_z(x,y) = \frac{B_0 L_0}{\pi} \cos(\frac{\pi x}{L_0}) \exp(- \frac{\pi y \sin \theta_0}{L_0}),
\end{eqnarray}
where $A_z$ is a component of the magnetic vector potential $\bm{A} = (A_x, A_y, A_z)$ of the force-free magnetic arcade, as we have $\nabla \times \bm{A} = \bm{B}$. The $A_z$ has a maximum value at the point $(x=0, y=0)$ and decreases with distance along any ray from this point. The $A_z(24\ \textrm{Mm},0)$ and $A_z(23\ \textrm{Mm}, 0)$ indicate two magnetic field lines in the $x-y$ plane. We note that the flare energy source $H_1$ is added only in the region given by Equation (\ref{q-aset}). The expression of $H_1$ is written as
\begin{equation}
H_1 = c_1 \frac{1}{\sqrt{\pi \lambda_l^2}} \exp [\frac{-(y-y_c)^2}{\lambda_l^2}] f(t) g(x), \label{q-H1}
\end{equation}
where $c_1$ denotes the total energy flux transported by energetic particles, $\lambda_l^2 = 0.05 \ \rm Mm^2$, and the height $y_c = 1.75 \ \rm Mm$ is located in the upper chromosphere. The function $f(t),$ which describes the temporal evolution of energy flux, is written as follows: 
\begin{equation}
f(t) = \frac{1}{\sqrt{\pi \lambda_t^2}} \exp [\frac{-(t-2 \lambda_t)^2}{\lambda_t^2}]. \label{q-ft}
\end{equation}
Obviously the energy flux has a temporal Gaussian distribution. The parameter $\lambda_t$ denotes the time scale of heating, and the maximum energy flux is achieved at $t = 2 \lambda_t$. The function $g(x),$ which indicates the asymmetry of energy deposition at both footpoints, is given by
\begin{equation}
g(x) = 
\begin{cases}
asym / (1 + asym),  & \quad x < 0 \\
1 / (1 + asym).  & \quad x > 0
\end{cases}
\label{q-as}
.\end{equation}
The parameter $asym$ denotes the ratios of energy deposited at the left footpoint ($x < 0$) to that at the right footpoint ($x > 0$). We set the ratio to 0.8 in most of our simulations, since asymmetric energy deposition is more likely to happen \citep{Aschwanden1999ApJ, Fang2016ApJ}.  Since the spatial distribution of the flare energy flux is static instead of dynamic, this method can not adapt to the change of density profile along the loop. The value of energy distributed per unit mass increases rapidly when a significant part of plasma leave this energy deposit area. As a result, the produced evaporation flows tend to have low densities and high speeds. However, this  did not influence our research too much because the trigger of KHI is determined by Alfv\'en Mach number of the flows rather than the speed of the flows. In our numerical study, we tended to produce evaporation flows with Alfv\'en Mach numbers comparable to the observational flows, rather than produce evaporation flows that have speeds comparable to the observational flows.

\section{Case study}

In this section, we study a high resolution case to investigate the temporal evolution of the flare loop, the radiative evolution of the flare loop, and the energy cascade process of turbulence. In this case, we adopted a spatial resolution of $19.5\ \rm km$ within and near the flare loop. The value of $c_1$ in Equation (\ref{q-H1}) is set to $ 1.288 \times 10^{13} \ \rm erg~ cm^{-2}$, which ensures that about $1 \times 10^{29} \ \rm erg$ of energy is deposited into the chromospheric footpoints during the simulation. The amount of energy equals the estimated total power of non-thermal electrons ($> 20 \ \rm keV$) in an M2.2 flare on 2003 June 10 \citep{Milligan2006ApJ}. We used a timescale $\lambda_t = 60 \ \rm s$ in Equation (\ref{q-ft}) to deposit $90\%$ of energy in 4 min and set the parameter $asym$ in Equation (\ref{q-as}) to 0.8 to achieve asymmetric footpoint heating.

\begin{figure*}[htbp]
\begin{center}
\includegraphics[width=\textwidth]{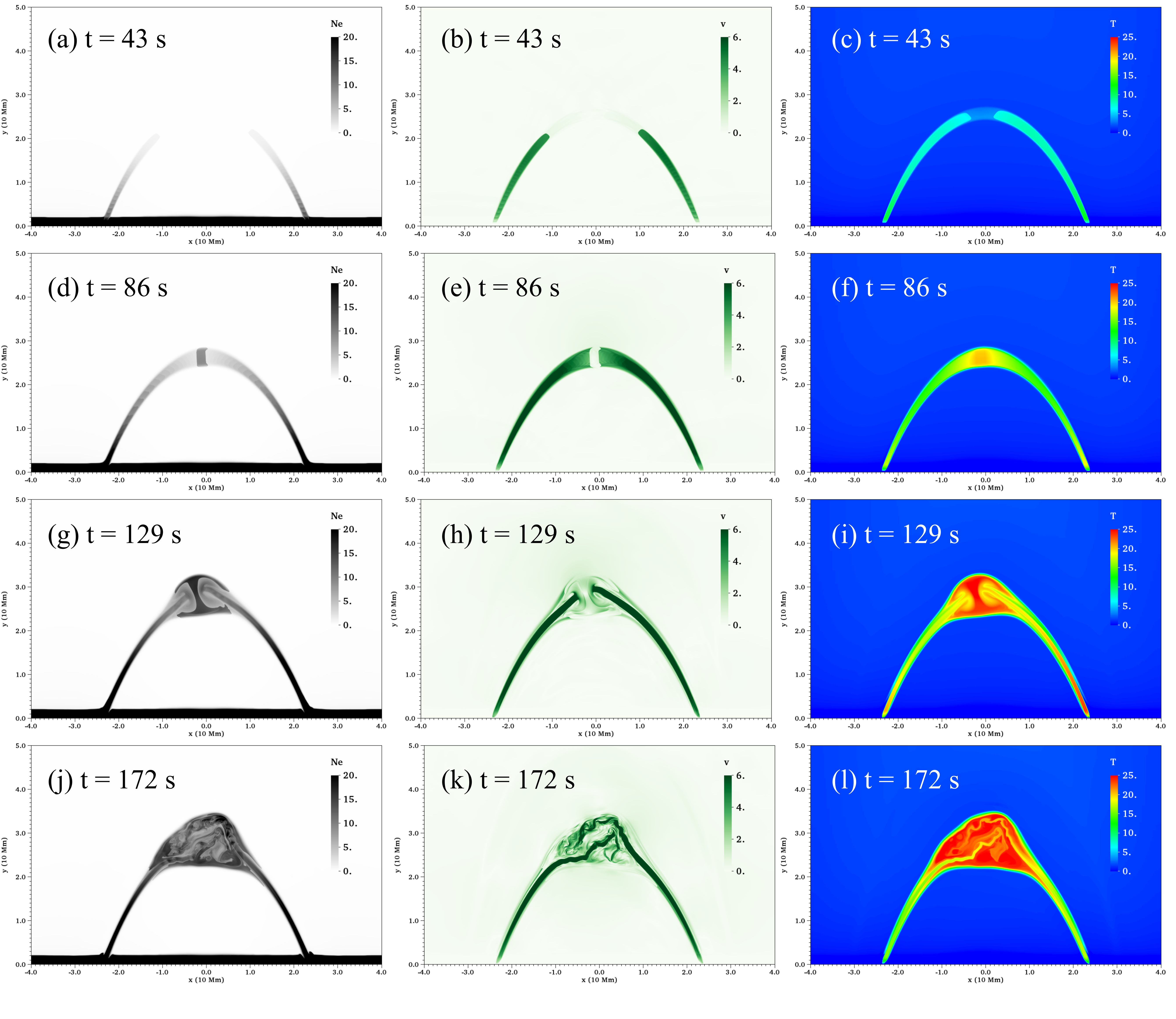}
\caption{Temporal evolution of number density (left column), velocity $v = \sqrt{v_x^2+v_y^2}$ (middle column), and temperature (right column) images at $t = 43, 86, 129,$ and $172\ \rm s$. The units of number density, velocity, and temperature are $10^9 \ \rm cm^{-3}$, $100 \ \rm km/s,$ and $\rm MK,$ respectively.}
\label{rhovT}
\end{center}
\end{figure*}

Temporal evolution of number density, speed, and temperature in this case are shown in Figure \ref{rhovT}. As a result of the sudden heating, the pressure in the chromospheric footpoints increases rapidly and two high speed evaporation flows are launched. Two hot, dense, and fast flows can be found at $t=43 \ \rm s$ in the top panels of Figure \ref{rhovT}. The densities of the evaporation flows are about $2 \times 10^{10} \ \rm cm^{-3}$ and the speeds are about $600 \ \rm km/s$. The Alfv\'en Mach number of the upflows seems to be comparable to that in the flare event reported by \citet{Tian2014ApJ}, where the electron density is of the order of $10^{11} \ \rm cm^{-3}$ and the blueshift is about $~ 260 \ \rm km/s$.
Even though the speed of the flows reaches as high as $600 \ \rm km/s$ in this case, no KHI is triggered near the boundary of the loop. The main reason is that the local Alfv\'en speed is also very high, as the Alfv\'en speed is about $600 \ \rm km/s$ in the inner loop and is higher than $4,000 \ \rm km/s$ outside the loop. When the dense flows meet each other near $t=86 \ \rm s$, a shock tube problem with two slow shocks is produced in the apex. The shock tube tends to expand along the magnetic field lines in the beginning, but this tendency is prevented quickly owing to the increase of number density and velocity in the evaporation flows that continue to impinge the bounding shocks. The increase of number density and velocity in the upstream drives the slow shock surfaces to move towards the middle, and the thermal pressure in the region between the two shock surfaces becomes higher and higher because of the plasma inflow. As a result, the high pressure forces the loop to expand vertically, as shown in panels (g)-(i) in Figure \ref{rhovT}. Because of the expansion, the strength of the local magnetic field becomes smaller and the Alfv\'en speed becomes lower in the loop apex. As another result of expansion, the evaporation flows shear with the dense plasma in the apex, and the KHI can be triggered since now both the conditions of high shear velocity and low Alfv\'en speed are satisfied. Thereafter, turbulence is produced by KHI, as shown in panels (j)-(l) of Figure \ref{rhovT}. Temperature fluctuations appear in the loop apex when turbulence has been produced. This temperature feature was pointed out in observational studies by \citet{Jakimiec1998A&A}.

We provide the synthesized AIA $131 \ \rm \AA$ emission and thermal SXR emission in $4-10 \ \rm keV$ for comparison with observations. The intensity of the AIA $131 \ \rm \AA$ line is given by 
\begin{equation}
I_{131} = G(T) n_e^2,
\end{equation}
where $G(T)$ is the contribution function provided by the CHIANTI atomic database \citep{DelZanna2015A&A}. The method to synthesize SXR emission is provided by \citet{Pinto2015A&A} and \citet{Fang2016ApJ}.  According to their method, the thermal X-ray photon flux at the photon energy $h \nu$ is 
\begin{equation}
I (h \nu, T) = I_0 \frac{EM}{h \nu \sqrt{k_{\textrm{b}} T}} g_{\textrm{ff}} (h \nu, T) \textrm{exp} (- \frac{h \nu}{k_{\textrm{b}} T}),
\end{equation}
where $g_{\textrm{ff}}$ is the Gaunt factor for free-free bremsstrahlung emission, which is
written as\begin{equation}
g_{\textrm{ff}} (h \nu, T) = 
\begin{cases}
1 , & h \nu \leq k_{\textrm{b}} T \\
(\frac{k_{\textrm{b}} T}{h \nu})^{0.4}, & h \nu > k_{\textrm{b}} T.
\end{cases}
\end{equation}
The coefficient $I_0 = 1.07 \times 10^{-42} $ if the photon flux $I (h \nu, T)$ is measured at a distance of $1 \ \rm AU$ and is expressed in units of $\rm cm^{-2} \ s^{-1} \ keV^{-1}$. For the emission measure of a finite volume of plasma $\textrm{EM} = n_e^2 V$, we used the volume of a cell $V = 19.5 \times 19.5 \times 19.5 \ \rm km^3$. We divided the energy range $4 < h \nu < 10 \ \rm keV$ equally into 60 pieces, and then estimated and added up the photon flux density in each piece  to calculate the total SXR flux density of the given range. The total SXR photon flux density can be described by $I_{\textrm{SXR}} = \sum I (h \nu, T) \Delta h \nu$. Integration of this emission flux along the light of sight was not performed, since our simulation data is 2D. Therefore, only the relative intensity of emission is meaningful.  

\begin{figure*}[htbp]
\begin{center}
\includegraphics[width=\linewidth]{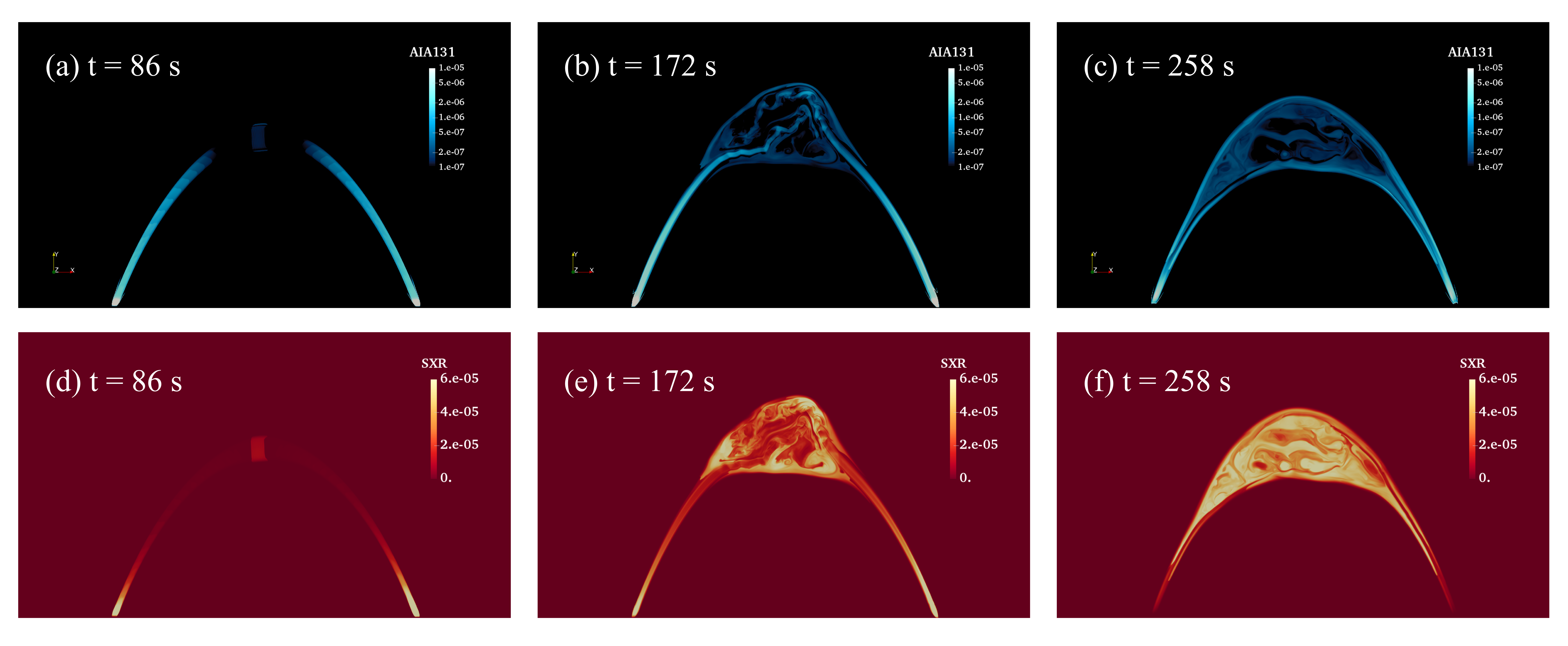}
\caption{Temporal evolution of synthesized AIA $131\ \rm \AA$ emission ($\rm DN \ cm^{-5} \ s^{-1} \ pix^{-1}$) and SXR emission ($\rm photon \ cm^{-2} \ s^{-1}$).}
\label{emission}
\end{center}
\end{figure*}

The evolution of SXR emission is demonstrated in Figure \ref{emission}. At $t = 86 \ \rm s$, high intensity emission only appears near the chromospheric footpoints, even though the evaporation flows have reached the apex. This is because the footpoints have higher plasma density and temperature ($\sim 17 \ \rm MK$). Later, the apex also shows high emission intensity at $t = 172 \ \rm s$, as the temperature of plasma increases to above $20 \ \rm MK$ via flow interaction. Thereafter, SXR emission disappears near the footpoints at $t = 258 \ \rm s$, as the heating is finished and the local temperature decreases to about $10 \ \rm MK$. Meanwhile, SXR emission at the apex is still strong as the temperature of plasma is still higher than $20 \ \rm MK$. The flare loop has a so-called fat and bright body accompanied by a so-called slim and less bright SXR emission down the legs. This shape is similar to the spider-like structure often observed in SXR images of flares \citep{Zhitnik2003MNRAS, Zhitnik2006SoSyR}. This spider-like shape is maintained until the end of the simulation ($t \approx 9 \ \rm min$). 

\begin{figure}[htbp]
\begin{center}
\includegraphics[width=1.0\linewidth]{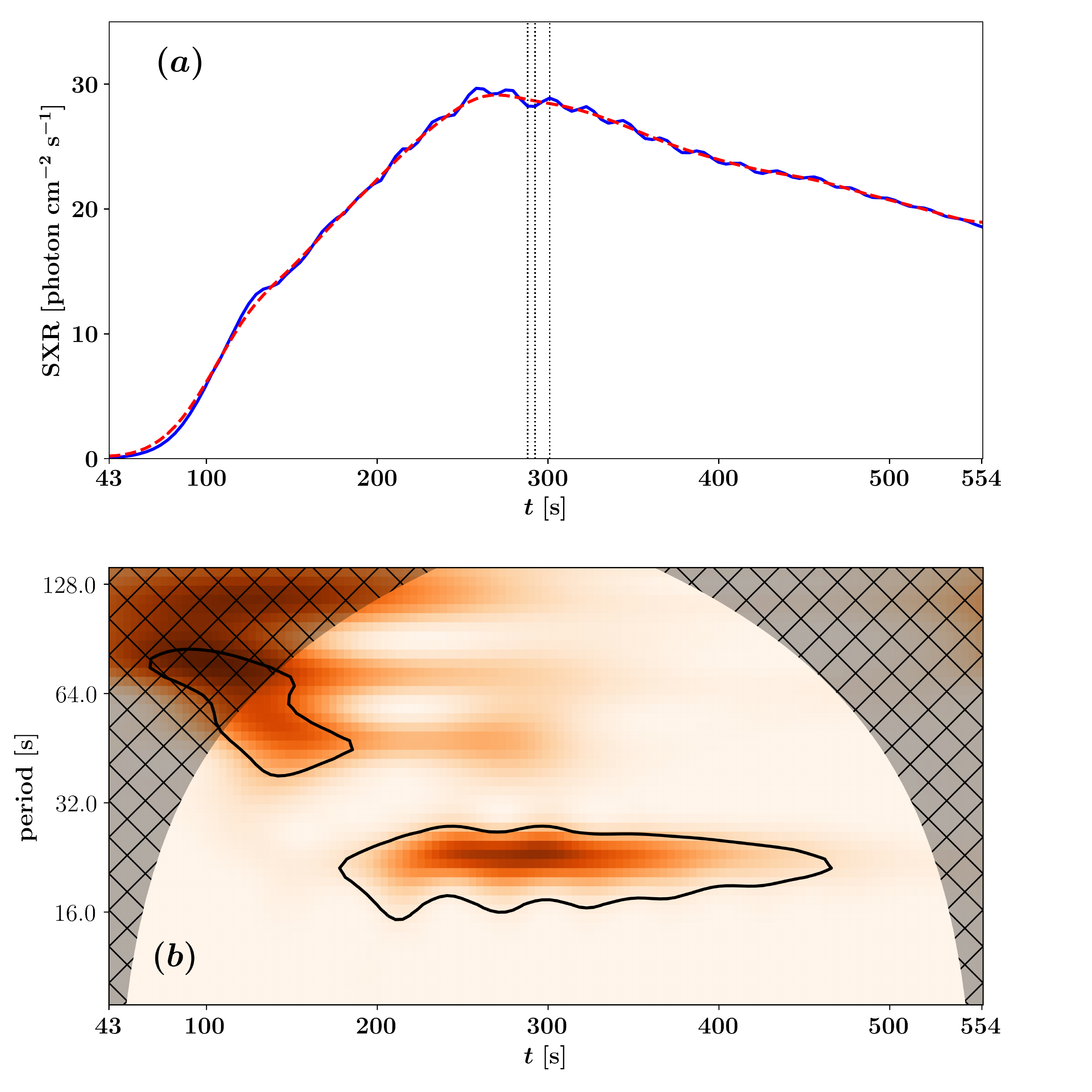}
\caption{ (a) Temporal evolution of the total flux of SXR emission (solid line) and the smooth estimate of the SXR emission (dashed line). (b) Wavelet analyses of the SXR emission. The black contour in panel (b) shows 95\% confidence level. The cross-hatched regions represent cone of influence regions, where the edge effect is important. Three dotted lines indicate the times $t = 288.1, 292.4$ and $301.0 \ \rm s,$ respectively.}
\label{SXR}
\end{center}
\end{figure}

\begin{figure}[htbp]
\begin{center}
\includegraphics[width=1.0\linewidth]{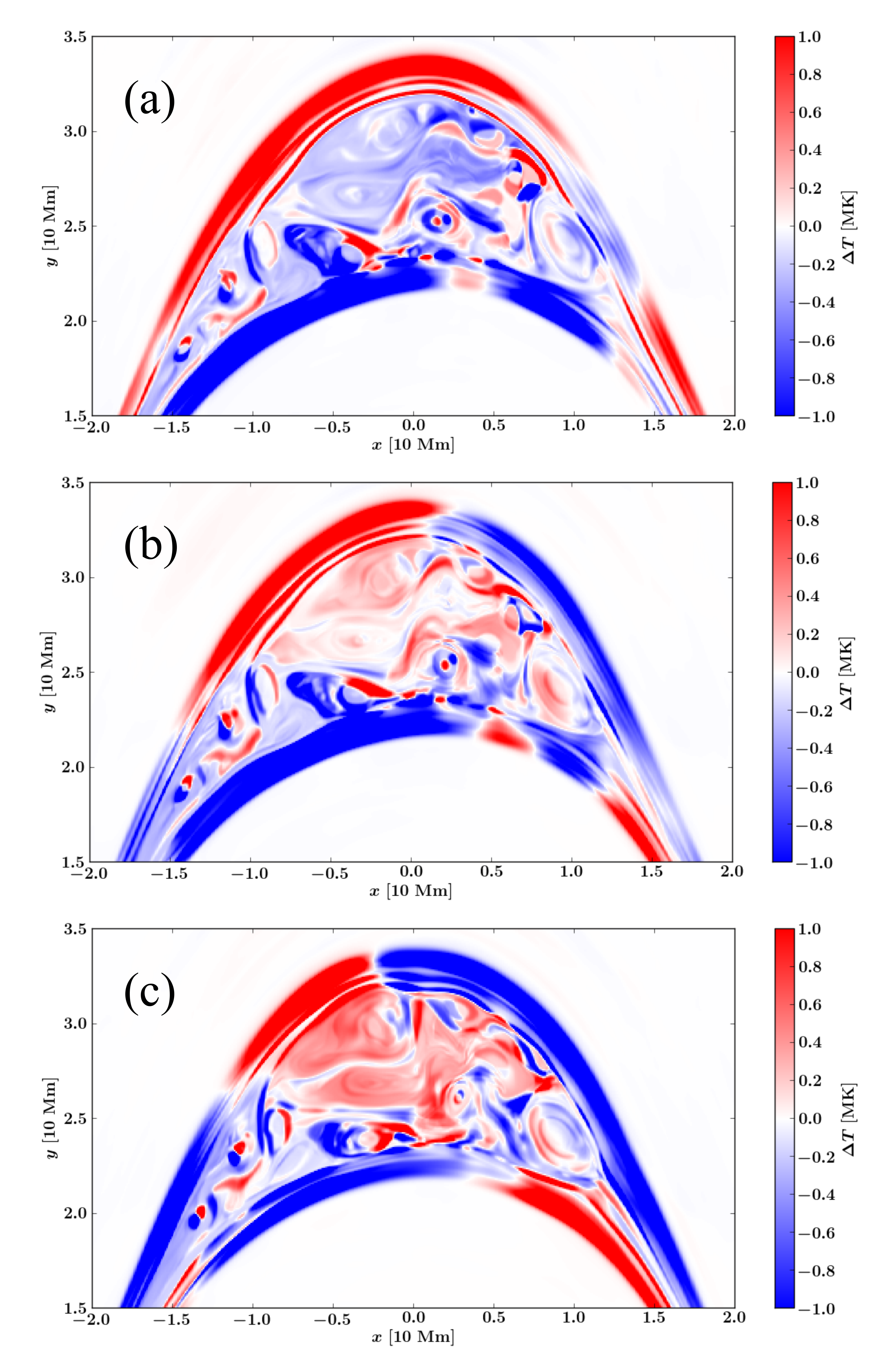}
\caption{Running difference analysis of temperature. (a): $T (t = 288.1~ \textrm{s}) - T (t = 283.8~ \textrm{s})$; (b):  $T(t = 292.4~ \textrm{s}) - T (t = 288.1~ \textrm{s})$; and (c): $T (t = 301.0~ \textrm{s}) - T (t = 296.7~ \textrm{s})$. The phases of oscillations at the times $t = 288.1, 292.4$ and $301.0 \ \rm s$ refer to Figure \ref{SXR}.}
\label{Te_diff}
\end{center}
\end{figure}

The total SXR flux of the flare heated loop at each time is calculated and plotted in panel (a) of Figure \ref{SXR}. The flux decreases to about $2/3$ of the maximum value when the simulation is finished at $558 \ \rm s$. Periodic signals are clearly demonstrated in the decreasing phase of the profiles. To investigate the periodic feature of the signals, wavelet analysis is performed and the results are demonstrated in panel (b). The original data is detrended before the wavelet analysis is performed. The method we perform in the detrending is given by 
\begin{equation}
s_d (t) = \frac{s_o (t)}{s_b (t)} - 1,
\end{equation}
where $s_d$ is the detrended data, $s_o$ is the original data and $s_b$ is the smooth estimate of $s_o$. The original data is smoothed with Gaussian filter. We used Morlet wavelet in the analysis and set the wave number to be $\omega_0 = 6$. The data before $t = 43 \ \rm s$ were not taken into account in the wavelet analysis, while the SXR flux in this interval is nearly 0. The results of wavelet analysis show that the signal has a clear single periodicity in the SXR decreasing phase, and the period is about $25 \ \rm s$. The oscillation has a long duration, which is more than five periods. To investigate the nature of the oscillations, we employed running difference analyses. The analyses show that temperature and density near the centre of the apex (i.e. in the turbulent region) also vary with a period of about $25 \ \rm s$. The results of the running difference analysis of temperature are demonstrated in Figure \ref{Te_diff}. The total SXR flux is temporally minimal at $t = 288.1 \ \rm s$ and is located at a peak at $t = 301.0 \ \rm s$. This figure clearly shows that the temperature near the centre of the apex decreases with time when the total SXR flux decreases with time (from $t = 283.8 \ \rm s$ to $t = 288.1 \ \rm s$), and the temperature increases with time when the total SXR flux increases with time (from $t = 288.1 \ \rm s$ to $t = 301.0 \ \rm s$). The variation of density is in phase with the variation of temperature as well. This indicates that the oscillations in the total SXR flux are compressional signals.

Oscillations in flares with periods of $\sim$ 1-60 s were interpreted as standing fast sausage modes \citep{Tian2016ApJ}. However, the oscillations in our simulation are difficult to interpret by this standard standing fast sausage wave model, which suggests that the phase speed of the waves is given by
\begin{equation}
C_p = 2 L / P\,,
\end{equation}
where $L$ is the loop length and $P$ is the period,
and the phase speed should be smaller than the external Alfv\'en speed \citep{Aschwanden2004psci}. For the oscillations in our simulation, we find $2 L / P \approx \rm160 \ Mm / 25\ s = 6400 \ km/s$. This value is much larger than the Alfv\'en speed inner the flare loop (lower than $1000 \ \rm km/s$) and close to the maximum Alfv\'en speed outside the flare loop; the outside Alfv\'en speed ranges from $4000 \ \rm km/s$ to $8000 \ \rm km/s$.
One probable interpretation is that the oscillations are standing fast sausage modes, where the waves are reflected before they arrive at the opposing footpoints. Instead, the waves are reflected near the apex because of the rapid change in loop radius. As a result, the wavelength is much shorter than $2 L$. We assume that the wavelength is $2 \times 30 \ \rm Mm = 60 \ \rm Mm$, while the size of the apex is about $20 - 30 \ \rm Mm$. In this assumption, the phase speed is $2400 \ \rm km/s$, which is lower than the Alfv\'en speed outside the apex and is a reasonable value of fast sausage wave phase speed.
 
Besides the $25 \ \rm s$ oscillations near the centre of the apex, another type of oscillations is found in the running difference analyses. These oscillations appear near the boundaries of the flare loop. As shown in Figure \ref{Te_diff}, negative $\Delta T$ propagates from right to left near the upper boundary, while positive $\Delta T$ propagates from right to left near the lower boundary. The positive or negative $\Delta T$ near the boundary is caused by the variation of the loop boundaries. The oscillation can be seen in animated views of the density variation, where the loop boundaries show a fairly coherent swaying motion. This is reflected in the temperature difference views shown in Figure \ref{Te_diff} where these coherent motions give clear correlated patterns in the temperature at the top and bottom edges of the loop (near its apex). These sideways displacements are characteristic for more kink-type perturbations.


When the density of the plasma and photon density are given, we can compare the role of ICS with that of bremsstrahlung in HXR emission. The equation (5) and (7) in appendix A of \citet{Krucker2008A&ARv} are expressions to estimated volume emissivity of ICS and bremsstrahlung, respectively. Here, we estimate the relative contribution of ICS versus bremsstrahlung with these two equations.
 For HXR photons of energy $\epsilon$, this ratio is written as
\begin{equation}
R (\epsilon) = \frac{3 \pi}{2 \alpha} \frac{n_{ph}}{n_i} (2\delta-1) Q(\delta) \left( \frac{\epsilon}{4 \epsilon_i} \right)^{(1-\delta)/2} \left( \frac{\epsilon}{m_e c^2} \right)^{\delta-1/2},
\end{equation}
where  $n_{ph}$ is the number density of mono-energetic photons that have energy $\epsilon_i$ (and are upscattered by ICS), $n_i$ is the ambient coronal number density, and $\delta$ is the spectral index of energetic electrons.  Furthermore, $\alpha$ is the fine-structure constant, and $m_e c^2$ is the rest energy of an electron. The dimensionless function $Q(\delta)$ is given by
\begin{equation}
Q (\delta) = \frac{2 (11 + 4 \delta + \delta^2)}{(1+\delta)(3+\delta)^2(5+\delta)} \,.
\end{equation}
The contribution of the inverse Compton process by scattering 2 eV photospheric photons, to 20 keV HXR photons, has been evaluated by \citet{Krucker2008A&ARv}. The authors suggest that the inverse Compton process may play a more important role than bremsstrahlung when the ambient coronal density is lower than $10^9 \ \rm cm^{-3}$. In our simulations, our plasma density is only $10^8 \ \rm cm^{-3}$ before the apex is filled with evaporation plasma, therefore their result indicates that this inverse Compton process may determine loop-top HXR emission only before evaporation flows reach the apex. The role of sustained loop-top turbulence, in higher density environments created by KHI interactions in which SXR photons are abundant, was not accounted for, however.

Since \citet{Fang2016ApJ} suggested that high energy electrons scatter SXR photons to HXR photons, and thereby contribute to the loop-top HXR source, we investigate this effect here. Firstly, we need to estimate the number density of SXR photons. According to our SXR synthesized results shown in Figure \ref{emission}, we can assume that the contribution of a single computational cell of volume $19.5 \times 19.5 \times 19.5 \ \rm km^3$ to the SXR flux measured at a distance of $R = \rm 1\ AU = 1.5 \ \times 10^{13} \ cm$ is of order $f_m = 5 \times 10^{-5} \rm \ photon \ cm^{-2} \ s^{-1}$. The volume of the loop-top SXR source can clearly be approximated with a sphere of radius 10 Mm, making the total emitting volume $V_l = (10\ \textrm{Mm})^3 4 \pi / 3 $ and the area of the loop-top  surface $S_l = 4 \pi (10\ \rm Mm)^2$. The total SXR flux is then estimated as $f_{t} = (V_l / V) f_m 4 \pi R^2$ and the SXR photon density is accordingly estimated as $n_{ph} = f_t / (S_l c) \approx 200 \ \rm cm^{-3}$, where $c$ is the light speed. To simplify the calculation, all SXR photons are assumed to have energy of 4 keV. The energetic electrons are assumed to have a hard spectrum $\delta = 2$, and the ion number density is obtained directly from our simulations and is of order $n_i = 10^{10} \ \rm cm^{-3}$. According to Equation (22), the relative contributions of scattering SXR photons to HXR photons and bremsstrahlung for the emission of 20 keV HXR photons is then $R\, (20\ \textrm{keV}) \approx 10^{-8}$. This implies that the contribution of the inverse Compton process to scattering SXR photons to HXR photons is very small for the generation of a loop-top HXR source. Still, we emphasize that the formulae used in this work do not account for the extra turbulence we find as an inevitable consequence of KHI. We now address these turbulence properties in what follows.

\begin{figure}[htbp]
\begin{center}
\includegraphics[width=\linewidth]{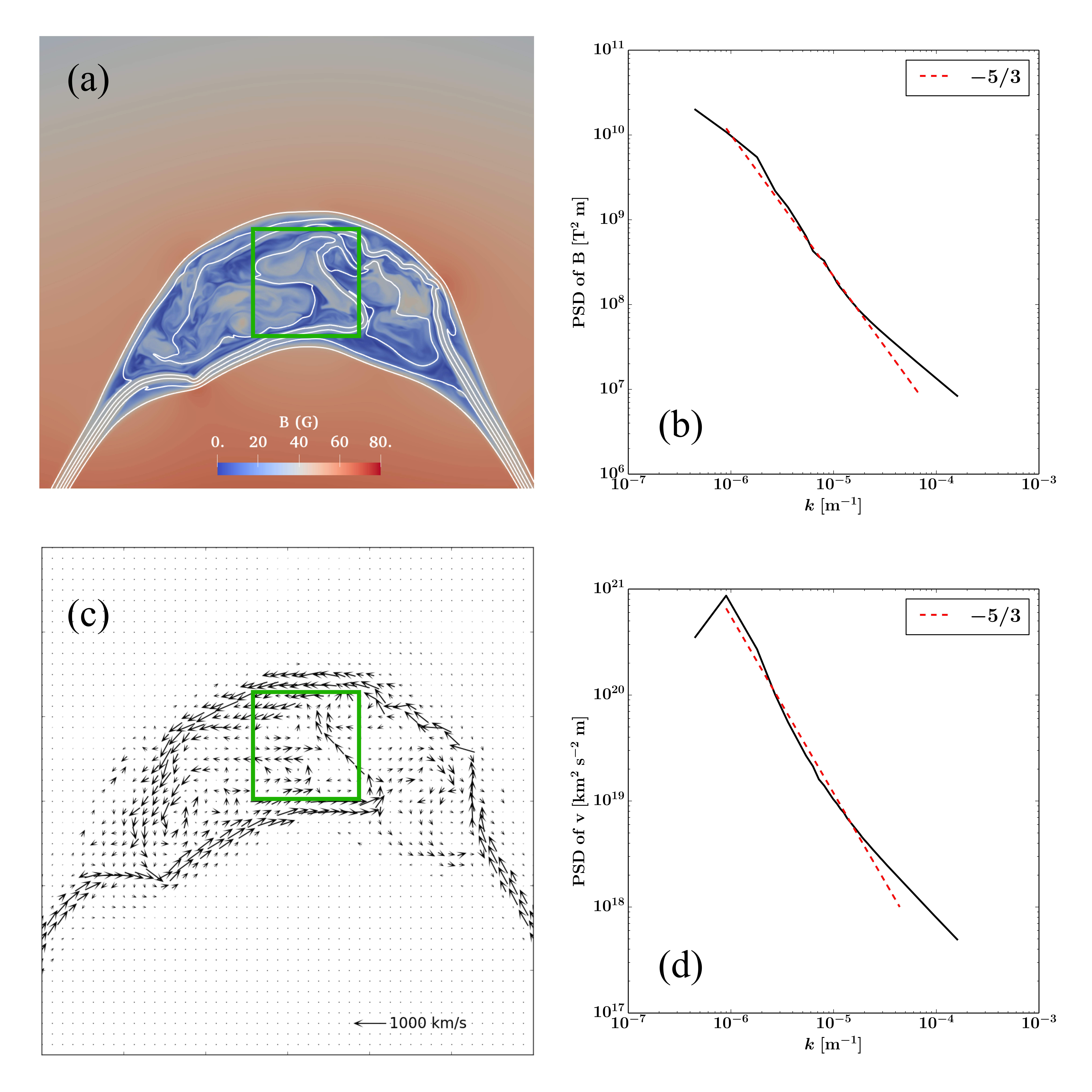}
\caption{(a) Magnetic field strength at $t = 215 \ \rm s$; (b) spectrum of magnetic field inside the green box; (c) flow field of velocity at $t = 215 \ \rm s$;  and (d) spectrum of velocity inside the green box. White lines in (a) denote magnetic field lines.}
\label{PSD}
\end{center}
\end{figure}

\begin{figure}[htbp]
\begin{center}
\includegraphics[width=1.0\linewidth]{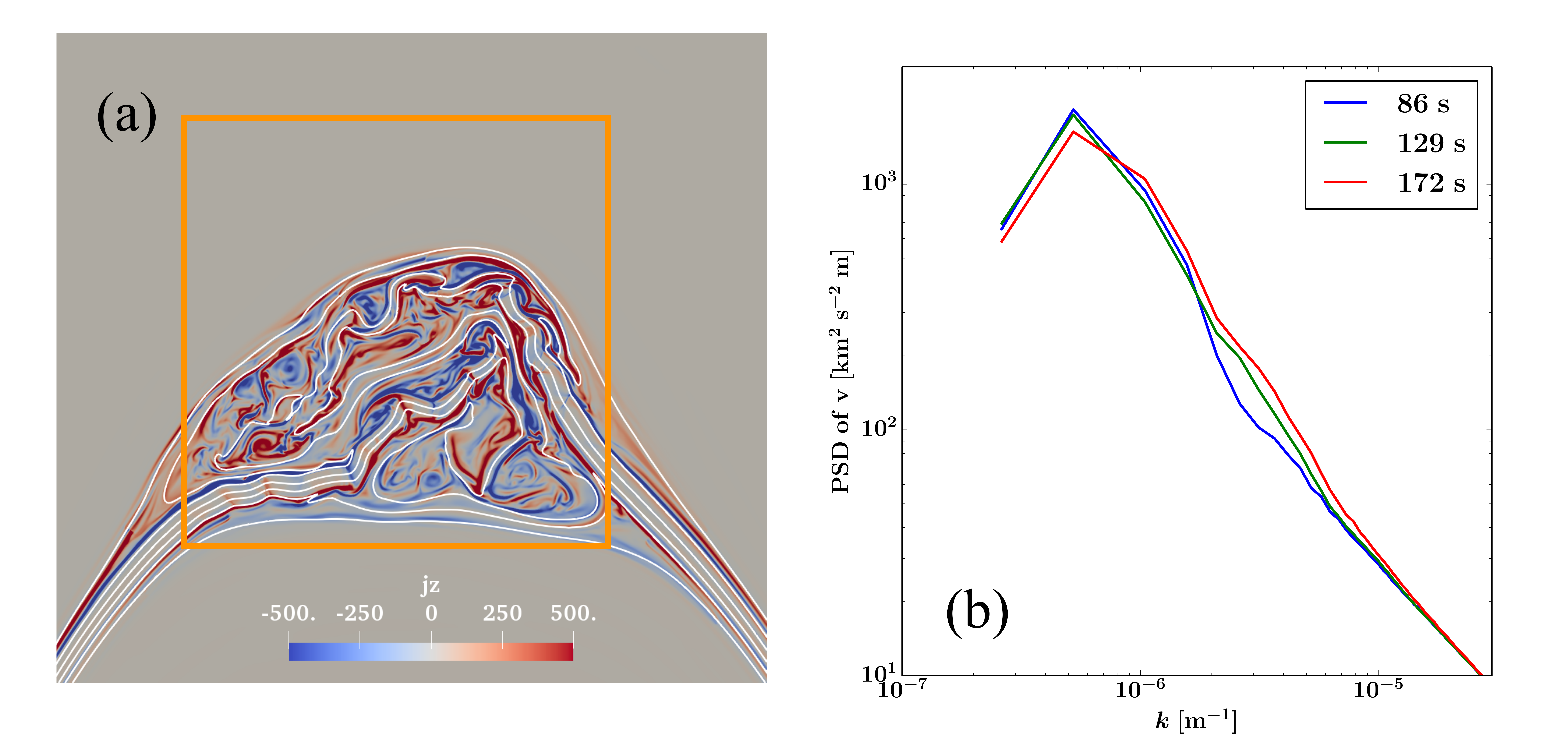}
\caption{(a) Out of plane current component $j_z$ at $t = 172 \ \rm s$; (b) temporal evolution of spectra. The region inside the orange box in panel (a) is selected to calculate the spectra.}
\label{psd_evl}
\end{center}
\end{figure}

Turbulence plays an important role in the new scenario presented by \citet{Fang2016ApJ}, as it is suggested to be an efficient accelerator for electrons and can trap high energy electrons in the loop top. Therefore, turbulence produced by KHI in the apex should be analysed. Figure \ref{PSD} shows the power spectral density (PSD) of velocity and magnetic field of the turbulence at $t=215 \ \rm s$, and the distribution of magnetic field and the flow field of velocity are also provided. A 2D region (green box in Figure \ref{PSD}) is selected to perform fast Fourier transform, and then the 2D results are organized to obtain 1D spectra. The 1D spectra are functions of wave number $k = \sqrt{k_x^2 + k_y^2}$, where $k_x$ and $k_y$ is the wave number in $x$ and $y$ direction respectively. Island structures and vortices can be found in the plots of magnetic field strength and velocity flow field, respectively. The spectral indexes are close to $-5/3$ in the spatial scale range of $10^{-6} < k < 10^{-5} \ \rm m^{-1} $. The cascade process of energy is clearly illustrated in Figure \ref{psd_evl}, which shows their temporal evolution.

To catch the energy cascade process, a larger region is selected to analyse the spectra (orange box in Figure \ref{psd_evl}a). Figure \ref{psd_evl}b provides the PSD of velocity at $t = 86, 129$ and $172 \ \rm s$. The value of PSD at the maximum wave number $k \approx  2 \times 10^{-4} \ \rm m^{-1}$ are set to one to compare the profiles. We find that the spectra coincide with each other very well in the range $k > 10^{-5} \ \rm m^{-1}$, which indicates that the energy cascade process stops at the scale $k \approx 10^{-5} \ \rm m^{-1}$. This may explain why the indexes of the spectra departs from $-5/3$ in the scale $k > 10^{-5} \ \rm m^{-1}$ in Figure \ref{PSD}. The energy cascades from large scale to small scale, as seen by comparing the three spectra: an enhancement can be found in scale $6 \times 10^{-6}  > k > 2 \times 10^{-6} \ \rm m^{-1}$ at $t = 129 \ \rm s,$ while compared with time $t = 86 \ \rm s$; spectra at $172 \ \rm s$ show an enhancement in $10^{-5} > k > 6 \times 10^{-6} \ \rm m^{-1}$ compared with $t = 129 \ \rm s$. The plasma states at $t = 86, 129,$ and $172 \ \rm s$ are shown on the full domain size in Figure \ref{rhovT}.

\begin{figure}[htbp]
\begin{center}
\includegraphics[width=\linewidth]{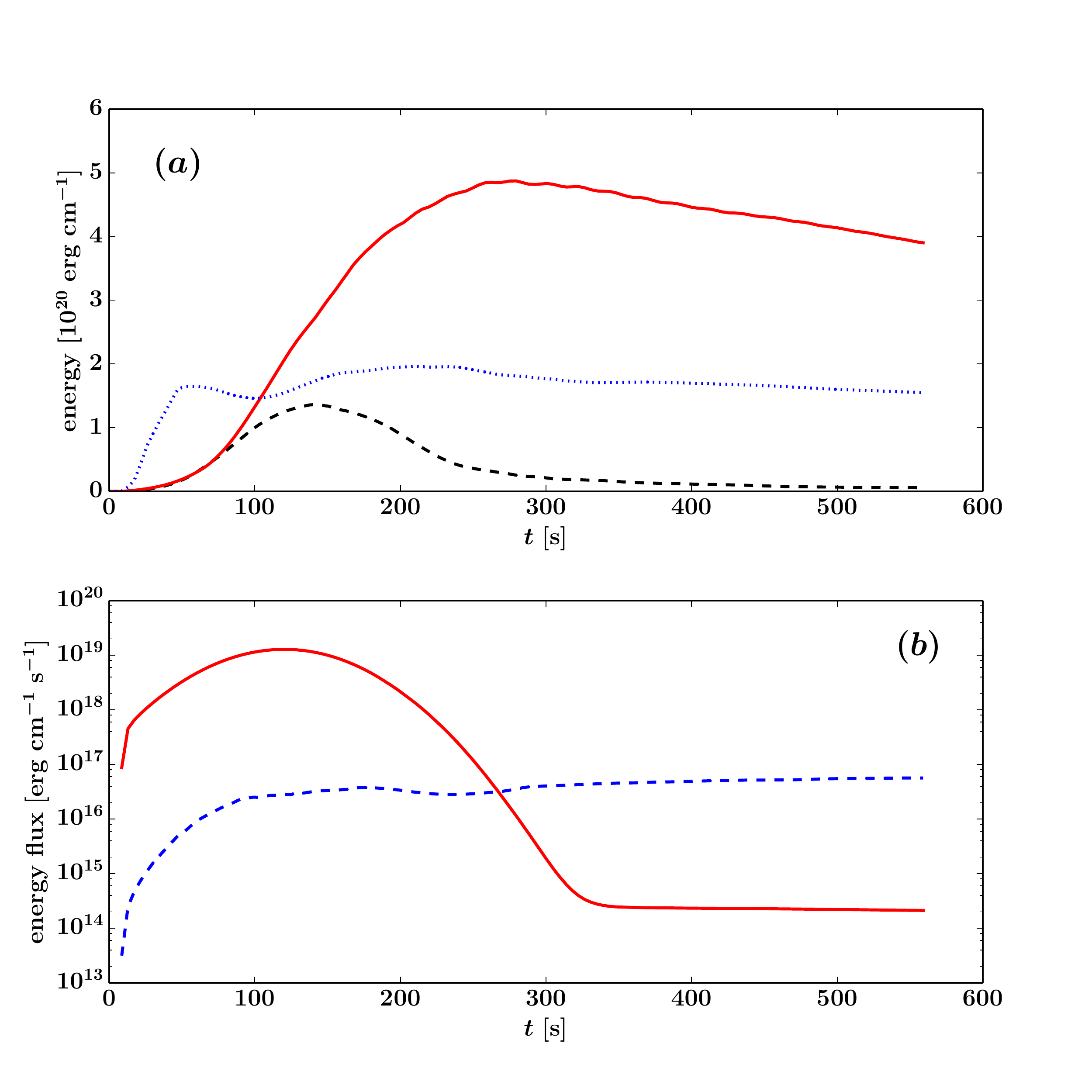}
\caption{(a) Temporal evolution of total kinetic energy (black dashed line), thermal energy (red solid line), and magnetic energy (blue dotted line) for hot plasma ($\rm T \geq 5 \ MK$); (b) evolution of background heating flux (red solid line), and radiative cooling (blue dashed line) for hot plasma.}
\label{energy}
\end{center}
\end{figure}

We are also interested in the energy transformation and heat-loss evolution of the flare loop. Therefore, the temporal evolutions of total kinetic energy, thermal energy and magnetic energy are calculated. We define the boundaries of the flare loop with the help of plasma temperature. The region where $T > 5 \ \rm MK$ is assumed to belong to the loop, since the plasma outside the loop has a temperature $T < 2 \ \rm MK$ and the temperature of plasma inside the loop is higher than $10 \ \rm MK$. The results are demonstrated in panel (a) of Figure \ref{energy}. The ratio of thermal pressure to magnetic pressure $\beta$ has a wide range in the loop apex. The value of $\beta$ ranges from 1 to 10 in the dark region in panel (e) of Figure \ref{emission}, while $\beta$ is as high as several tens in the bright region in this panel. In contrast, $\beta$ is about $5 \ \times 10^{-4}$ in the apex and has an order of 1 in the chromospheric footpoints when energy has not been deposited. Before the evaporation flows collide with each other at $t \approx 75 \ \rm s$, the kinetic energy flow has the same value as the thermal energy flow. After the flows meet each other in the apex, compression causes kinetic energy to transform quickly into thermal energy. The evidence can be found in Figure \ref{rhovT}, where the plasma in the apex is rapidly heated to about $ 20 \ \rm MK$ from about $10 \ \rm MK$ when the flows collide with each other. The plasma in the loop remains hot for a long time after most of the energy is deposited in the footpoints, since the radiative cooling is less efficient compared with the heating (Figure \ref{energy}) in the hot loop. The temperature of the loop is about $20 \ \rm MK$ when the simulation is finished at $t = 558 \ \rm s$, and weak turbulence can still be found in the apex. The plasma temperature in our simulation is comparable to that in the flare observations. \citet{Nitta2001ApJ}  investigated the loop-top temperatures in 36 flare events, and the temperatures in their study are in the range $18 - 23.5 \ \rm MK$.

\section{Parameter surveys}

In this section, we investigate what may influence the trigger of KHI and the generation of turbulence. The driving source of the evaporation flows, i.e. the chromospheric energy deposition process, is expected to have great influence on the dynamics of the flare loop. Therefore, we explore parameters of the energy deposition that are considered to be influential: the amount of energy, timescale of deposition, and asymmetry of deposition. 

The influence of the energy deposition timescale is investigated with two cases in this paper. Both cases have a resolution of $39 \ \rm km$. Two new cases listed in this work are referred to as case 2 and case 3, respectively, while the case in section 3 is referred to as case 1. Case 2 has the same setting as case 1 except for the resolution. Case 3 has $\lambda_t = 90 \ \rm s$ to let $90 \%$ of the energy be deposited in about 6 minutes, while $\lambda_t = 60 \ \rm s$ in case 2. Hence, the amounts of energy are the same but case 2 has a shorter timescale and stronger instantaneous energy flux than case 3.

\begin{figure}[htbp]
\begin{center}
\includegraphics[width=\linewidth]{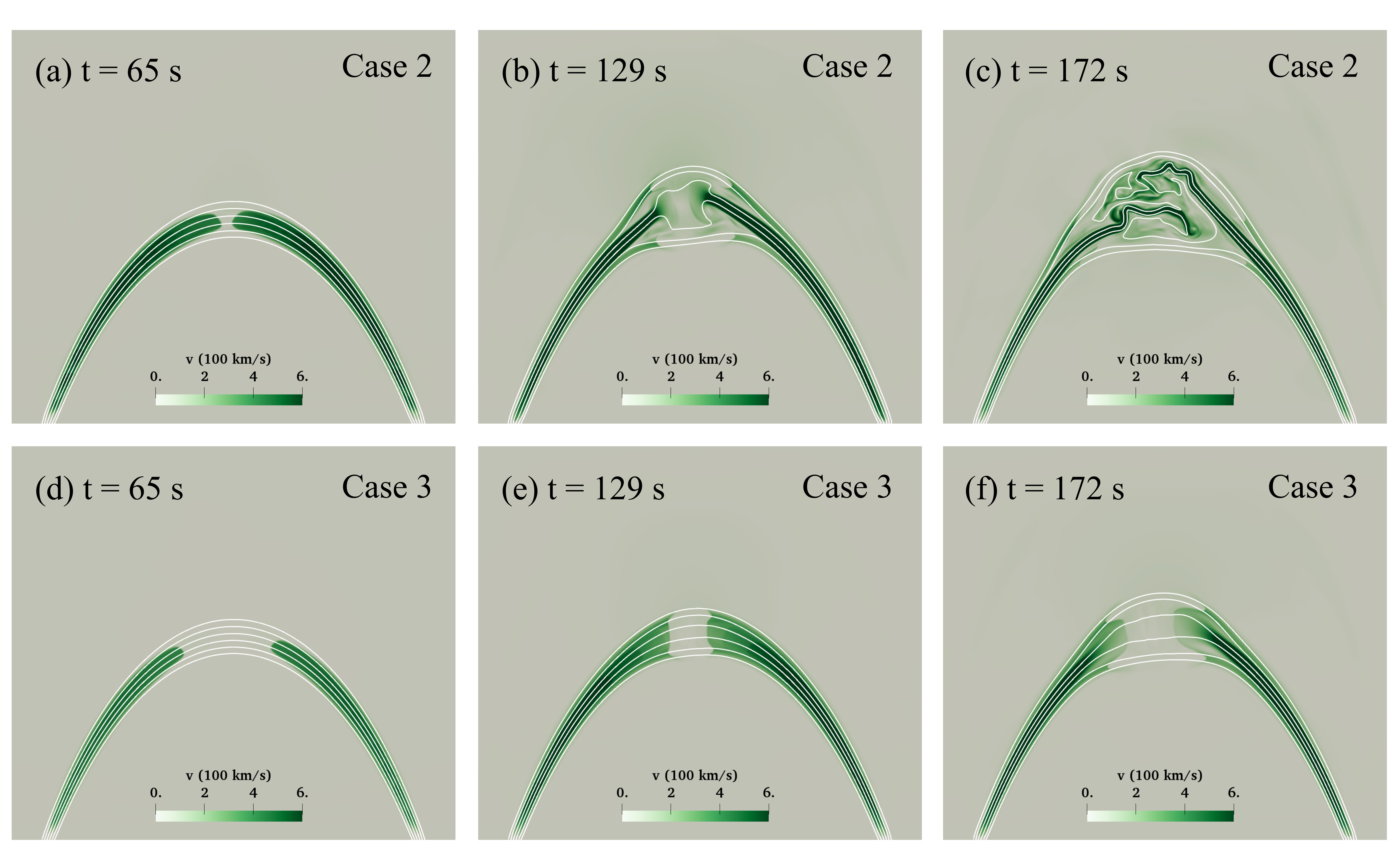}
\caption{Evolution of velocity for case 2 in which $\lambda_t = 60 \ \rm s$ (a, b, c) and case 3 in which $\lambda_t = 90 \ \rm s$ (d, e, f).}
\label{dtime}
\end{center}
\end{figure}

\begin{figure}[htbp]
\begin{center}
\includegraphics[width=1.0\linewidth]{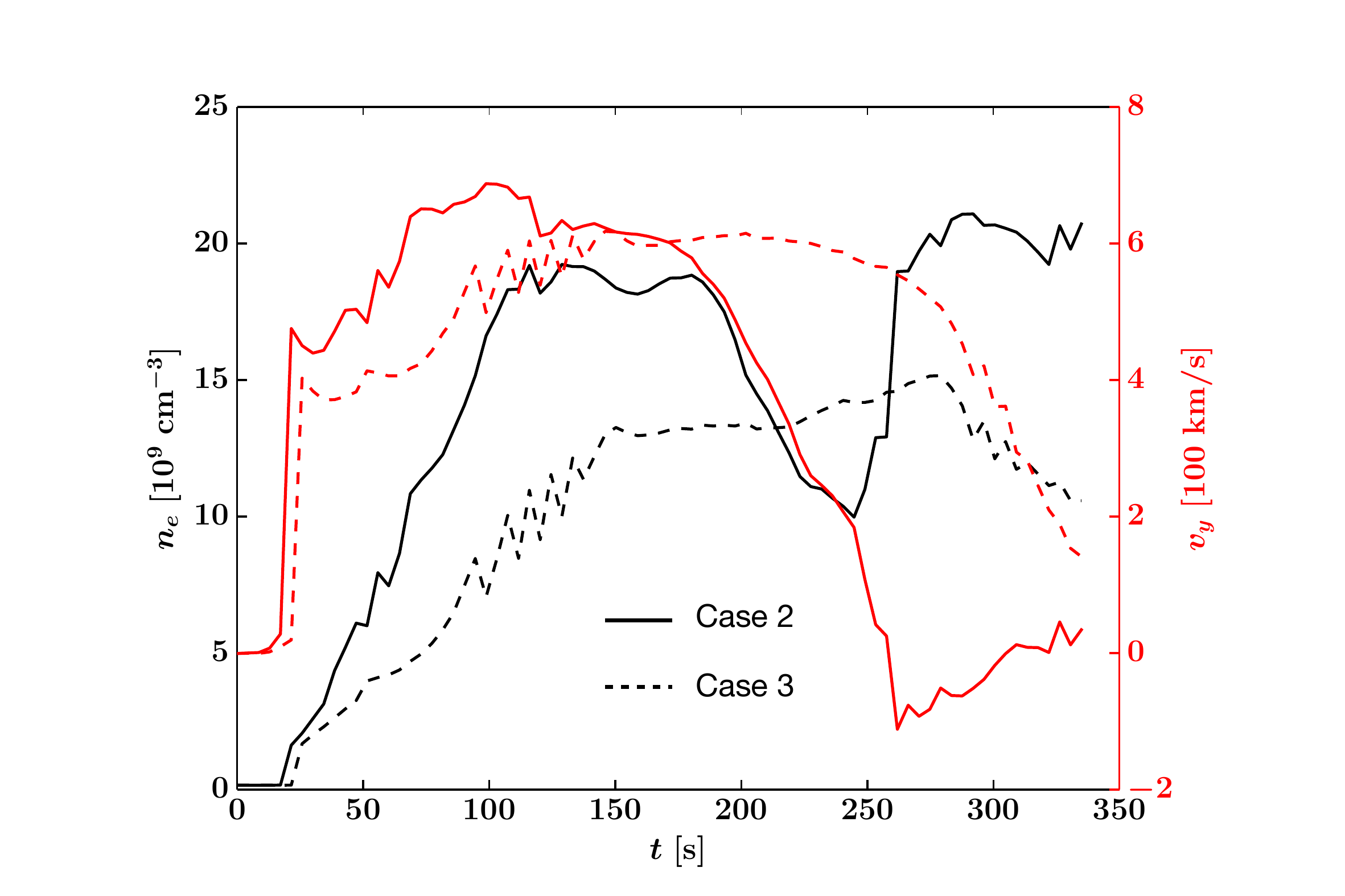}
\caption{Temporal evolution at the centre of the right loop leg at the height of $h = 10 \ \rm Mm$ of density (black lines) and velocity (red lines) for case 2 with higher energy flux and case 3 (dashed) with lower energy flux.}
\label{rhov}
\end{center}
\end{figure}

The temporal evolution of the velocity of case 2 is shown in Figure \ref{dtime}a-\ref{dtime}c, and that of case 3 is in Figure \ref{dtime}d-\ref{dtime}f. In case 2, turbulence is produced owing to KHI and vortices can be found in Figure \ref{dtime}c. In contrast, no turbulence is produced and no shear flows can be found in case 3. To study the difference between case 2 and case 3, the parameters of the evaporation flows are analysed. The temporal evolution of density and velocity at the axis of the right loop at a height of $h = 10 \ \rm Mm$ is shown in Figure \ref{rhov}. Figure \ref{rhov} indicates a positive correlation between density or velocity of the flows and the energy flux of footpoint heating. The energy flux is higher in case 2 compared with case 3, since the  same amount of energy is deposited in a shorter time. Moreover, the density or velocity of the flow increase faster before a maximum value is achieved in case 2. The flows collide near the apex at $t \approx 70 \ \rm s$ in case 2 and at $t \approx 80 \ \rm s$ in case 3. The maximum speed of the flows  is about $700 \ \rm km/s$ in case 2 and about $600 \ \rm km/s$ in case 3 at the measured point.

As demonstrated in Figure \ref{rhovT}, the flows collide with each other in the apex and two slow shocks are generated before the KHI is triggered. From the Rankine-Hugoniot equations we know that the propagating velocity of a shock is determined by the velocity and density and pressure in the upstream and downstream. For case 2, the shock surface near the axis propagates towards the downstream since the velocity and density in the upstream increase quickly. At the same time, the shock fronts near the loop boundaries propagate upstream to maintain balances of pressure and mass transport because the velocity and density of flows near the boundary are much smaller than those near the axis. As a result, shear flows are generated in the loop apex and KHI can be triggered (Figure \ref{dtime}b). In contrast, the density and velocity of upstream in case 3 increase slowly, and therefore the shock front near the axis propagates upstream. Because the shock fronts near the boundaries propagate towards upstream as well, no shear flows can be found in the apex and KHI cannot be triggered. In addition, stronger energy flux leads to larger Alfv\'en Mach numbers.  The Alfv\'en Mach number at the axis of the right loop at a height of $h = 10 \ \rm Mm$ at $t=151 \ \rm s$ is about $1.5$ in case 2 and about $0.9$ in case 3. This means that KHI is easier triggered in case 2 even if shear flows exist in the apex in both cases. For the same reason, the more energy is deposited the easier KHI is triggered.

\begin{figure}[htbp]
\begin{center}
\includegraphics[width=\linewidth]{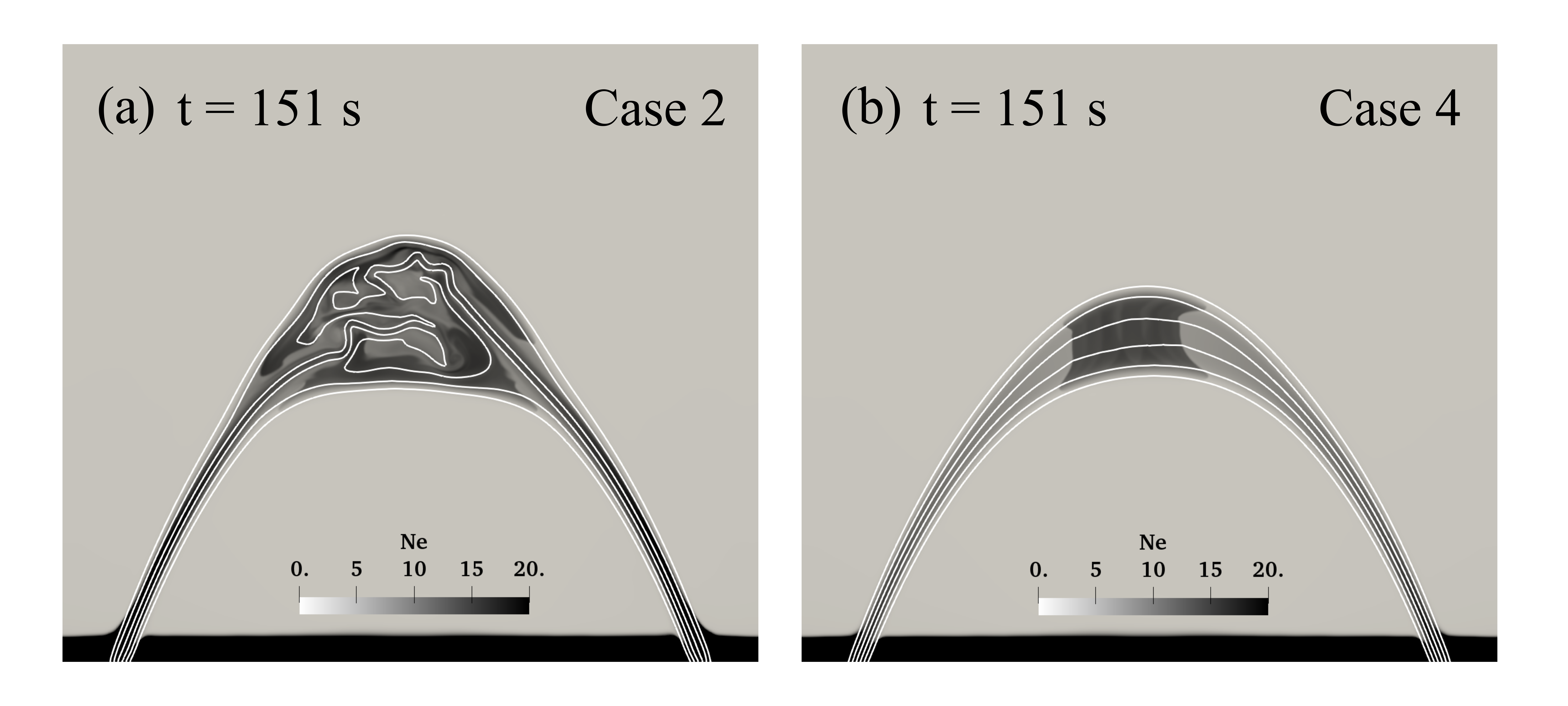}
\caption{(a) Number density at $t = 151 \ \rm s$ for case 2 with more energy deposited at the footpoints; (b) density at $t = 151 \ \rm s$ for case 4 with less energy deposited. The unit of number density is $10^{9} \ \rm cm^{-3}$.}
\label{denergy}
\end{center}
\end{figure}

A new case (case 4) with a resolution of $39 \ \rm km$ is simulated and compared with case 2 to study the influence of the amount of deposited energy. The amount of energy deposited into the footpoints in case 4 is set to $1/3$ of that in case 2 with the same timescale of energy deposition. Figure \ref{denergy}a and \ref{denergy}b show the distribution of density at $t = 151 \ \rm s$ of case 2 and case 4, respectively. In case 2, the density and velocity of the evaporation flows increase quickly, since the energy flux deposited into the footpoints increases quickly. As a result, the flows push both shocks to the middle and high speed flows can go into the loop apex. Therefore, KHI can be triggered and turbulence can be produced. In contrast, evaporation flows fail to go into the apex in case 4, since the density and velocity of the flows increase slowly. Consequently, the flows fail to produce turbulence in this case.

\begin{figure}[htbp]
\begin{center}
\includegraphics[width=\linewidth]{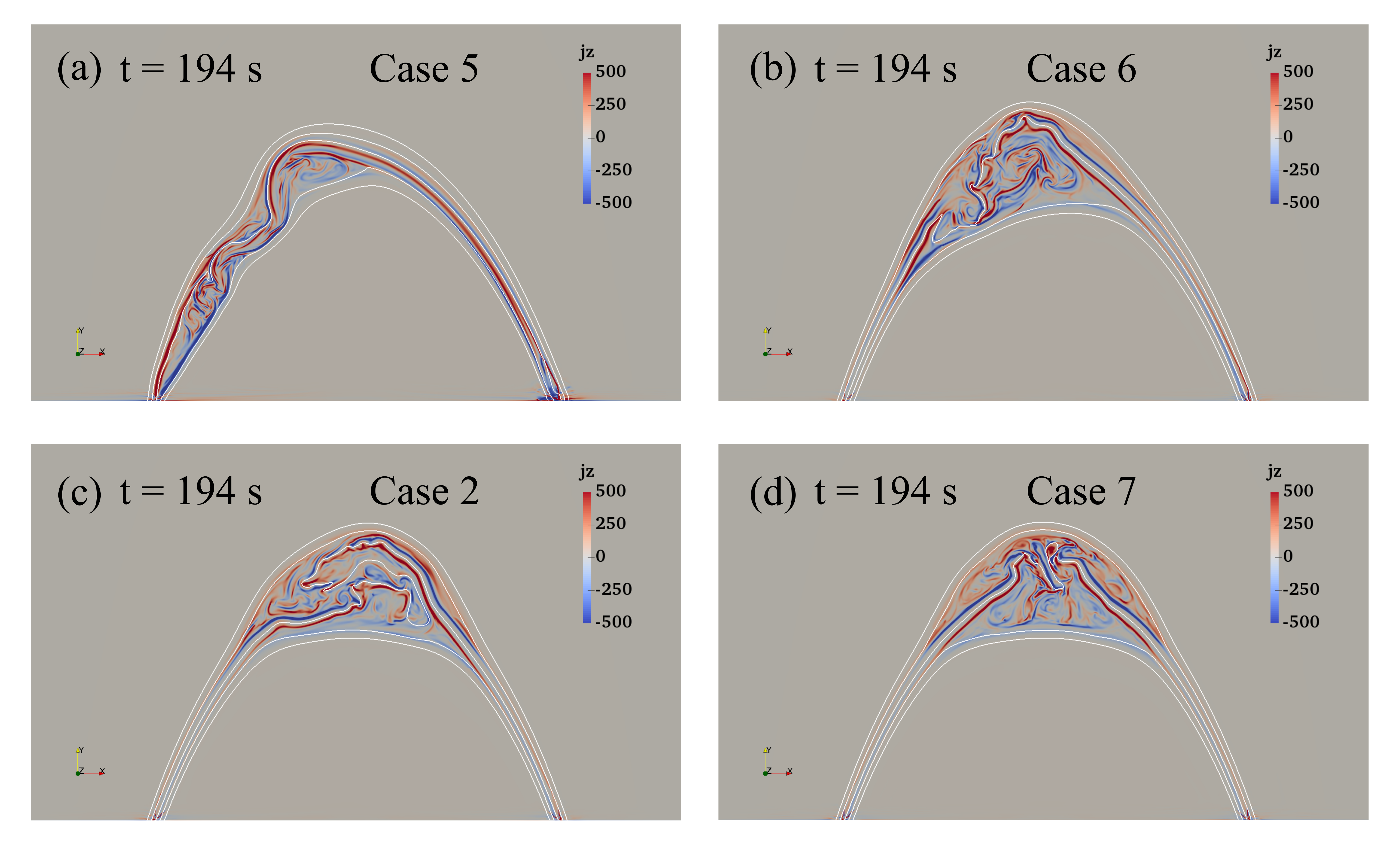}
\caption{Out of plane current component $j_z$ at $t = 194 \ \rm s$ for four cases with different energy deposition asymmetries. The asymmetric parameters $asym$ in panels (a), (b), (c), and (d) are 0, 0.4, 0.8 and 1, respectively.}
\label{asym}
\end{center}
\end{figure}

In the observation of solar flares, the HXR emission often exhibits asymmetric features \citep{Aschwanden1999ApJ}. The asymmetry of HXR emission is likely caused by the asymmetric energy deposition at two footpoints. 
Therefore, it is meaningful to study the influence of asymmetry of energy deposition on the KHI and turbulence. Three new cases (case 5, 6, and 7) are simulated and compared with case 2 to investigate the influence in our work. In these four cases, the amounts of energy deposited in the loop and the timescales of energy deposition are the same, but the ratios of energy deposited at the left footpoint to the right foot-point are different. The ratios in cases 5, 6, and 7 are set to $0$, $0.4$ and $1.0,$ respectively, while that in case 2 is $0.8$. The corresponding simulation results of the out of plane current component $j_z$ at $t = 194 \ \rm s$ are demonstrated in Figure \ref{asym}. The resolution of the cases is $39 \ \rm km$. The vortical structures in each panel provide the information about the location where the KHI is triggered and turbulence is produced. Figure \ref{asym} denotes that the location where turbulence is produced is determined by the asymmetry of energy deposition. KHI is always triggered near the location where the interaction of the flows happen. Since the velocity of evaporation flow has a positive correlation with the energy flux deposited in the footpoint, an asymmetry of energy deposition leads to an asymmetry of flow speed. Consequently, KHI tends to be triggered near the apex when the ratio is close to $1$ and tends to be triggered away from the apex when the ratio is close to $0$, as shown in Figure \ref{asym}.  We can still observe asymmetric vortex breaking when the heating is symmetric owing to numerical symmetry breaking and inaccuracy as shown in panel (d).

\begin{figure}[htbp]
\begin{center}
\includegraphics[width=\linewidth]{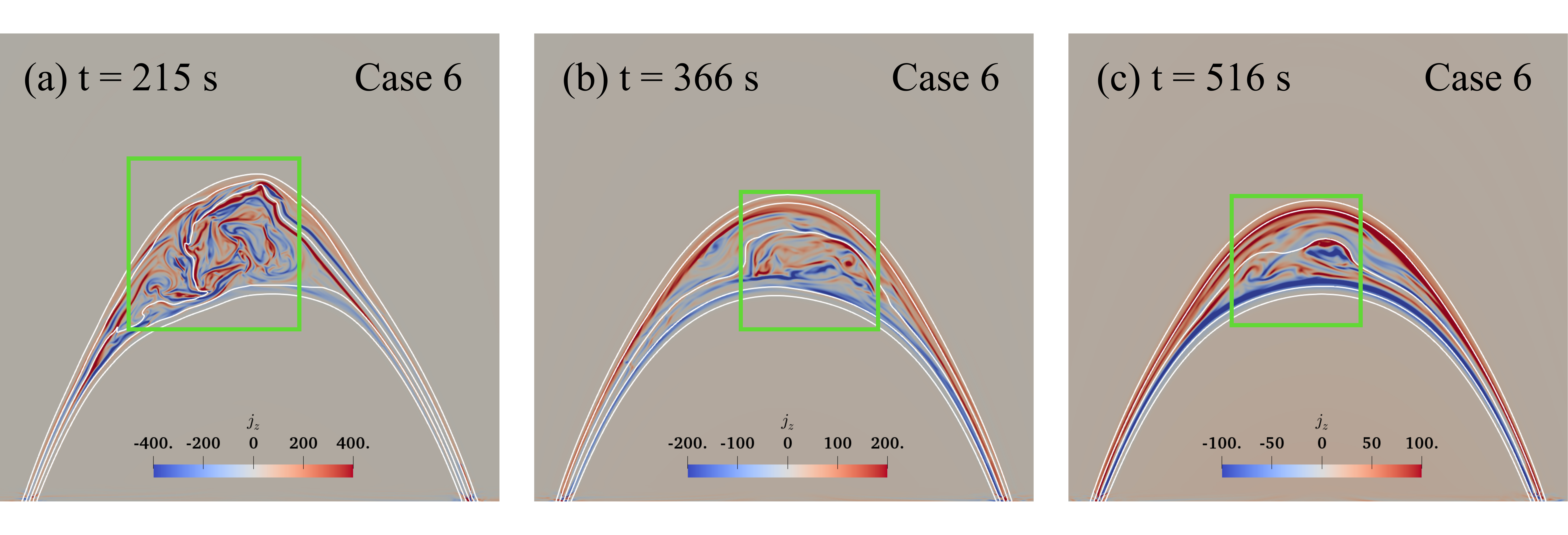}
\caption{Temporal evolution of out of plane current component $j_z$ for case 6 where $asym = 0.4$. Turbulence is in the left of apex at $ t = 215 \ \rm s$, and then appear in the right of apex at $t = 366 \ \rm s$. Finally, turbulence can be found in the left of the apex at $t = 516 \ \rm s$.}
\label{motion}
\end{center}
\end{figure}

When the turbulence is produced away from the apex because of asymmetric footpoint heating, the vortical structures tend to move back and forth near the apex (Figure \ref{motion}). The motion of the vortical structures seems to have a period of about $300 \ \rm s$. The periodic motion is probably a signal of standing slow-mode wave. The length of the loop is about $80 \ \rm Mm$ and the sound speed of $20 \ \rm MK$ plasma is about $500 \ \rm km/s$. The period of a standing slow wave should be about $320 \ \rm s$, which is close to the period of the motion in our simulations. \citet{Fang2015ApJ} and \citet{Mandal2016ApJ} have demonstrated that footpoint heating can generate slow waves and these reflect back and forth in the loop. 

\begin{figure}[htbp]
\begin{center}
\includegraphics[width=\linewidth]{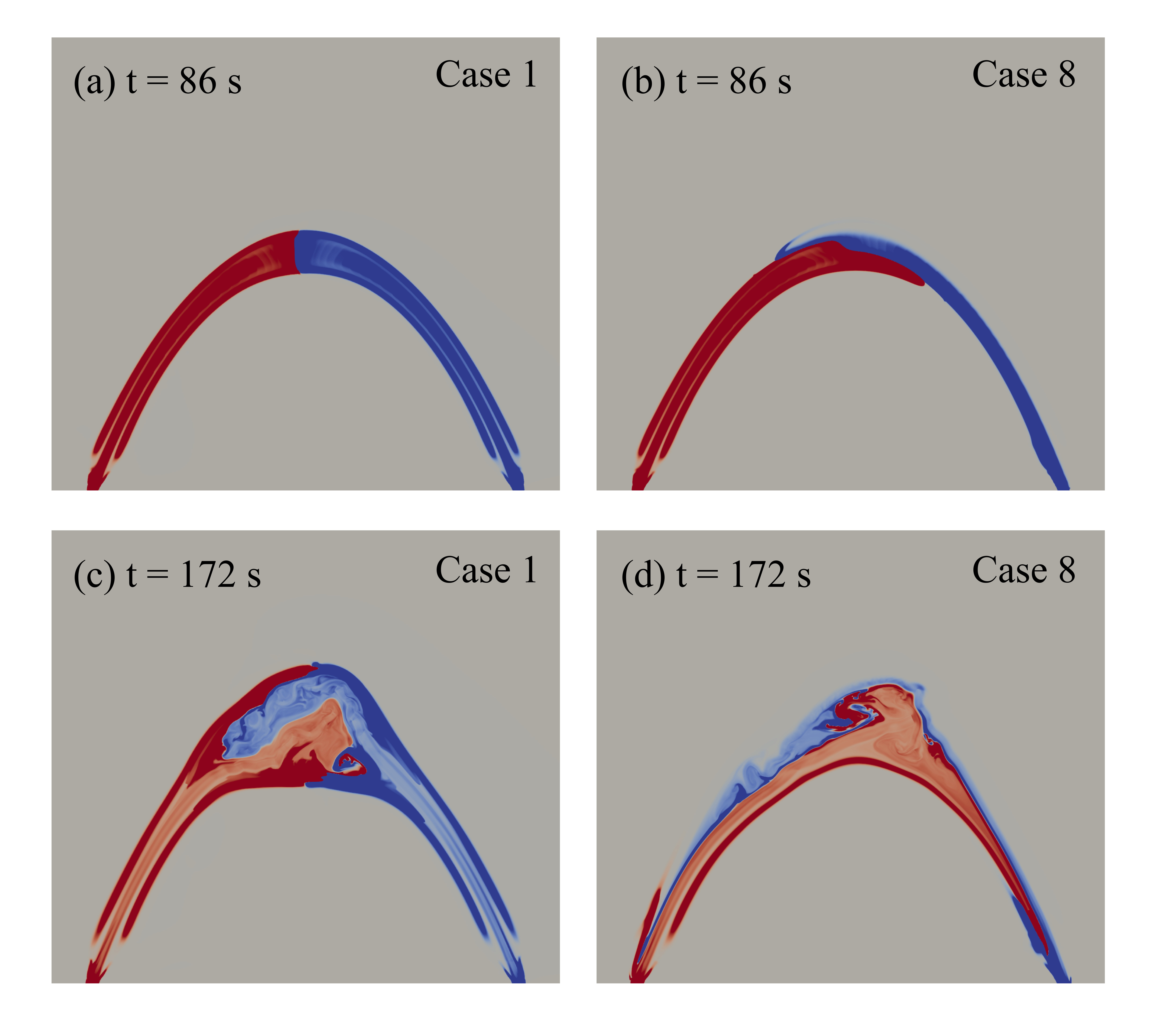}
\caption{Trigger of KHI for two ways the evaporation flows interact with each other. Red denotes the plasma from left footpoint, while blue denotes plasma from the right footpoint. Panels (a) and (c): flows from two footpoints collide with each other; (b)\&(d): flows shear with each other. }
\label{shear}
\end{center}
\end{figure}

A case (case 8) in which two evaporation flows shear with each other is also studied. This case has a high resolution of $19.5 \ \rm km$, the same as the resolution of case 1 in section 3. In this new case, energy is deposited at various magnetic field lines to ensure that the evaporation flows shear each other rather than collide with each other. All other cases have the same set of magnetic field lines that are affected at both left and right footpoint, as given by equation~(\ref{q-aset}). In
Figure \ref{shear}, the results of this new case are compared to that of case 1 where the flows collide with each other. Vortical structures can be found in both cases, which denotes that KHI can be triggered in both cases. However, more vortices can be found in the situation that the flows collide with each other. The collision of the evaporation flows leads to an expansion of the flare loop, a decrease of magnetic field strength and a decrease of local Alfv\'en speed. Therefore, KHI is more easily triggered in the situation in which the flows collide with each other. The interface of two evaporation flows is wider in the shearing case (case 8), which leads to KHI turbulence appearing on a long part of the loop. In contrast, two evaporation flows make contact with each other only in the apex in the collision case (case), which leads to a more local site of KHI and turbulence development.

\section{Summary}

We numerically study a new scenario proposed by \citet{Fang2016ApJ} concerning the origin of the observed turbulence in flare loops and generation of loop-top HXR emission. According to this new scenario, KHI can be triggered when chromospheric evaporation flows from two footpoints that meet and interact with each other in the flare loop apex, and turbulence can be produced by the KHI. The turbulence can act as a trapping region and an efficient accelerator to provide energetic electrons. The accelerated electrons can scatter SXR photons to HXR photons via the inverse Compton process to generate the loop-top HXR source, and add to the HXR bremsstrahlung, although the plasma densities are too high according to parametrized emission models \citet{Krucker2008A&ARv}. We focus on the trigger of KHI and the generation of turbulence. In our numerical study, energy is deposited into the chromospheric footpoints of a coronal loop and evaporation flows are produced owing to the sudden heating. The interaction of the flows, trigger of KHI, and generation of turbulence in the apex are investigated. The temporal evolution of a flare loop, radiative evolution of the flare loop, and energy cascade process of turbulence are studied with a high spatial resolution case. Thereafter, a parameter survey is performed to investigate what may influence the trigger of KHI and the generation of turbulence. The main results are summarized in the following: 

(1) KHI can be triggered when the evaporation flows interact with each other and turbulence can be produced by the KHI in our simulations. 

(2) Thermal SXR emission is synthesized and SXR sources are found in the results. The SXR emission appear in the footpoint at first, and then appear in the apex. The footpoint sources disappear when the footpoint heating is finished, but the loop-top source is maintained for several minutes.

(3) The spectral indexes of velocity and magnetic field are close to $-5/3$ in the turbulent region. The energy cascade process in the turbulence is clearly demonstrated in the spectra. 

(4) When the evaporation flows are produced, the kinetic energy and thermal energy are very close. Kinetic energy is quickly transformed into thermal energy due to compression after the flows collide with each other at $t \approx 75 \ \rm s$ and the loop becomes very hot ($T > 20 \ \rm MK$).

(5) The loop is still very hot ($T > 20 \ \rm MK$) when the simulation is finished at $t \approx 10 \ \rm minutes$ owing to inefficient radiative cooling. 

(6) The KHI is easily triggered when the density and velocity of the evaporation flows increase rapidly. The KHI turbulence is more easily produced when the amount of energy deposited into the footpoints is higher or the timescale of energy deposition is shorter.

(7) The location of the turbulence is determined by the ratio of energy deposited at two footpoints. The site of KHI and turbulence development moves away from loop apex when the energy deposited into one footpoint is much more than that deposited into the other footpoint.

(8) The conditions in which evaporation flows collide with other and the flows shear with each other are studied, and KHI turbulence can be produced in both conditions.

(9) Abundant waves are produced in the flare loops, including compressional modulations in the turbulent zone with a period of about $25 \ \rm s$ and standing slow mode wave with a period of about  $300 \ \rm s$.

In our simulations, the upward velocity of the flows can reach $700 \ \rm km/s$ (see Figure \ref{rhov}). Such a high velocity is not necessary to trigger KHI. Whether or not KHI can be triggered is determined by the Alfv\'en Mach number, which depends on both velocity and density. In the observations, it does not seem easy for evaporation flows to achieve a velocity of $700 \ \rm km/s$, but the density of plasma tends to be higher. For example, the electron density of the observed evaporation flows is found to be of the order of $10^{11} \ \rm cm^{-3}$ in \citet{Tian2014ApJ}, which is several times what we found in our simulations. Meanwhile, the observed evaporation flows are reported to have a blueshift of $\sim 260 \ \rm km/s$. The Alfv\'en Mach number of the observed evaporation flows is comparable to that in our simulations if the strengths of magnetic field are of the same order. Therefore, we suggest that KHI is not difficult to trigger in solar flares. 

Future work is required to study the same scenario in 3D setting to get a clear insight into orientational line of sight effects. A flare loop that contains a reconnection site needs to be adopted to incorporate the KHI effects in a real flare scenario. A more dynamic energy deposit method that distributes flare energy base on the density profile along the loop should be adopted. Test particles need to be added into the simulation to study the effect and physics of particle acceleration and trapping in KHI turbulence. The emission fluxes of HXR generated via ICS and bremsstrahlung can be estimated with the distribution of the test particles. Such calculations can quantify the hardness ($\delta$) of the electron spectrum. The inverse Compton mechanism may contribute to loop-top HXR emission before evaporation flows go into the apex. The contribution is then from scattering  photospheric lower energy photons to HXR photons rather than from scattering SXR photons to HXR photons. Bremsstrahlung dominates the HXR emission when the apex is filled with evaporation plasma, according to parametrized models that do not incorporate the turbulence effects. 
Future work must quantify the roles of the inverse Compton and bremsstrahlung mechanisms in the generation of loop-top HXR sources by comparing the HXR emissions produced via these mechanisms.

\begin{acknowledgements}

This work is supported by the Chinese scholarship council (CSC) and by project GOA/2015-014 (2014-2018 KU Leuven).  We would like to thank Tom Van Doorsselaere, Kirit Makwana, Hui Tian, and Zheming Guo for stimulating discussions. The simulations were conducted on the VSC (Flemish Supercomputer Center funded by Hercules foundation and Flemish government).

\end{acknowledgements}

\bibliographystyle{aa}
\bibliography{references}

\end{document}